\documentclass[longauth]{aa}

\usepackage{graphicx}
\usepackage{natbib}
\usepackage{scalerel}
\usepackage{overpic}
\usepackage[table]{xcolor}

\usepackage{txfonts}
\usepackage[pdfencoding=auto,psdextra]{hyperref}
\hypersetup{
    colorlinks=true,
    linkcolor=blue,
    filecolor=magenta,      
    urlcolor=blue,
    citecolor=blue
}
\makeatletter
\renewcommand*\aa@pageof{, page \thepage{} of \pageref*{LastPage}}
\makeatother

\usepackage[utf8]{inputenc}

\usepackage[switch, modulo]{lineno}
\linenumbers

\usepackage{euclid}

\newcommand{\lensmc}{\textsc{LensMC}\xspace}
\newcommand{\healpix}{\texttt{HEALPix}\xspace}
\usepackage{enumitem}
\setlist[enumerate]{itemsep=2pt, topsep=2pt}
\usepackage{hyperref}
\usepackage{lastpage}

\defcitealias{EP-Congedo}{C24}

\begin{document}
%
%

\title{\Euclid: Quick Data Release (Q1) -- LensMC shear measurement catalogue for cluster lensing science%
\thanks{This paper is published on behalf of the Euclid Consortium.}%
}


\newcommand{\orcid}[1]{} 
\author{G.~Congedo\orcid{0000-0003-2508-0046}\thanks{\email{giuseppe.congedo@ed.ac.uk}}\inst{\ref{aff1}}
\and M.~Sereno\orcid{0000-0003-0302-0325}\inst{\ref{aff2},\ref{aff3}}
\and H.~Miyatake\orcid{0000-0001-7964-9766}\inst{\ref{aff4},\ref{aff5},\ref{aff6}}
\and S.~Guerrini\orcid{0009-0004-3655-4870}\inst{\ref{aff7}}
\and M.~Kilbinger\orcid{0000-0001-9513-7138}\inst{\ref{aff8}}
\and M.~Radovich\orcid{0000-0002-3585-866X}\inst{\ref{aff9}}
\and H.~Jansen\orcid{0009-0002-1332-7742}\inst{\ref{aff10}}
\and F.~Kleinebreil\orcid{0009-0008-0116-2307}\inst{\ref{aff10}}
\and T.~Schrabback\orcid{0000-0002-6987-7834}\inst{\ref{aff10}}
\and A.~N.~Taylor\inst{\ref{aff1}}
\and B.~Altieri\orcid{0000-0003-3936-0284}\inst{\ref{aff11}}
\and L.~Amendola\orcid{0000-0002-0835-233X}\inst{\ref{aff12},\ref{aff13}}
\and S.~Andreon\orcid{0000-0002-2041-8784}\inst{\ref{aff14}}
\and N.~Auricchio\orcid{0000-0003-4444-8651}\inst{\ref{aff2}}
\and C.~Baccigalupi\orcid{0000-0002-8211-1630}\inst{\ref{aff15},\ref{aff16},\ref{aff17},\ref{aff18}}
\and M.~Baldi\orcid{0000-0003-4145-1943}\inst{\ref{aff19},\ref{aff2},\ref{aff3}}
\and S.~Bardelli\orcid{0000-0002-8900-0298}\inst{\ref{aff2}}
\and P.~Battaglia\orcid{0000-0002-7337-5909}\inst{\ref{aff2}}
\and A.~Biviano\orcid{0000-0002-0857-0732}\inst{\ref{aff16},\ref{aff15}}
\and R.~P.~Blake\inst{\ref{aff1}}
\and E.~Branchini\orcid{0000-0002-0808-6908}\inst{\ref{aff20},\ref{aff21},\ref{aff14}}
\and M.~Brescia\orcid{0000-0001-9506-5680}\inst{\ref{aff22},\ref{aff23}}
\and S.~Camera\orcid{0000-0003-3399-3574}\inst{\ref{aff24},\ref{aff25},\ref{aff26}}
\and V.~Capobianco\orcid{0000-0002-3309-7692}\inst{\ref{aff26}}
\and C.~Carbone\orcid{0000-0003-0125-3563}\inst{\ref{aff27}}
\and V.~F.~Cardone\inst{\ref{aff28},\ref{aff29}}
\and J.~Carretero\orcid{0000-0002-3130-0204}\inst{\ref{aff30},\ref{aff31}}
\and M.~Castellano\orcid{0000-0001-9875-8263}\inst{\ref{aff28}}
\and G.~Castignani\orcid{0000-0001-6831-0687}\inst{\ref{aff2}}
\and S.~Cavuoti\orcid{0000-0002-3787-4196}\inst{\ref{aff23},\ref{aff32}}
\and A.~Cimatti\inst{\ref{aff33}}
\and C.~Colodro-Conde\inst{\ref{aff34}}
\and L.~Conversi\orcid{0000-0002-6710-8476}\inst{\ref{aff35},\ref{aff11}}
\and Y.~Copin\orcid{0000-0002-5317-7518}\inst{\ref{aff36}}
\and F.~Courbin\orcid{0000-0003-0758-6510}\inst{\ref{aff37},\ref{aff38},\ref{aff39}}
\and H.~M.~Courtois\orcid{0000-0003-0509-1776}\inst{\ref{aff40}}
\and M.~Cropper\orcid{0000-0003-4571-9468}\inst{\ref{aff41}}
\and H.~Degaudenzi\orcid{0000-0002-5887-6799}\inst{\ref{aff42}}
\and G.~De~Lucia\orcid{0000-0002-6220-9104}\inst{\ref{aff16}}
\and C.~Dolding\orcid{0009-0003-7199-6108}\inst{\ref{aff41}}
\and H.~Dole\orcid{0000-0002-9767-3839}\inst{\ref{aff43}}
\and F.~Dubath\orcid{0000-0002-6533-2810}\inst{\ref{aff42}}
\and X.~Dupac\inst{\ref{aff11}}
\and M.~Farina\orcid{0000-0002-3089-7846}\inst{\ref{aff44}}
\and R.~Farinelli\inst{\ref{aff2}}
\and S.~Ferriol\inst{\ref{aff36}}
\and F.~Finelli\orcid{0000-0002-6694-3269}\inst{\ref{aff2},\ref{aff45}}
\and P.~Fosalba\orcid{0000-0002-1510-5214}\inst{\ref{aff46},\ref{aff47}}
\and S.~Fotopoulou\orcid{0000-0002-9686-254X}\inst{\ref{aff48}}
\and M.~Frailis\orcid{0000-0002-7400-2135}\inst{\ref{aff16}}
\and M.~Fumana\orcid{0000-0001-6787-5950}\inst{\ref{aff27}}
\and L.~Gabarra\orcid{0000-0002-8486-8856}\inst{\ref{aff49}}
\and S.~Galeotta\orcid{0000-0002-3748-5115}\inst{\ref{aff16}}
\and K.~George\orcid{0000-0002-1734-8455}\inst{\ref{aff50}}
\and B.~Gillis\orcid{0000-0002-4478-1270}\inst{\ref{aff1}}
\and C.~Giocoli\orcid{0000-0002-9590-7961}\inst{\ref{aff2},\ref{aff3}}
\and J.~Gracia-Carpio\orcid{0000-0003-4689-3134}\inst{\ref{aff51}}
\and A.~Grazian\orcid{0000-0002-5688-0663}\inst{\ref{aff9}}
\and F.~Grupp\inst{\ref{aff51},\ref{aff52}}
\and S.~Hemmati\orcid{0000-0003-2226-5395}\inst{\ref{aff53}}
\and M.~S.~Holliman\inst{\ref{aff1}}
\and W.~Holmes\orcid{0009-0007-8554-4646}\inst{\ref{aff54}}
\and I.~M.~Hook\orcid{0000-0002-2960-978X}\inst{\ref{aff55}}
\and F.~Hormuth\inst{\ref{aff56}}
\and A.~Hornstrup\orcid{0000-0002-3363-0936}\inst{\ref{aff57},\ref{aff58}}
\and K.~Jahnke\orcid{0000-0003-3804-2137}\inst{\ref{aff59}}
\and M.~Jhabvala\inst{\ref{aff60}}
\and S.~Kermiche\orcid{0000-0002-0302-5735}\inst{\ref{aff61}}
\and B.~Kubik\orcid{0009-0006-5823-4880}\inst{\ref{aff36}}
\and M.~K\"ummel\orcid{0000-0003-2791-2117}\inst{\ref{aff52}}
\and M.~Kunz\orcid{0000-0002-3052-7394}\inst{\ref{aff62}}
\and H.~Kurki-Suonio\orcid{0000-0002-4618-3063}\inst{\ref{aff63},\ref{aff64}}
\and A.~M.~C.~Le~Brun\orcid{0000-0002-0936-4594}\inst{\ref{aff65}}
\and S.~Ligori\orcid{0000-0003-4172-4606}\inst{\ref{aff26}}
\and P.~B.~Lilje\orcid{0000-0003-4324-7794}\inst{\ref{aff66}}
\and V.~Lindholm\orcid{0000-0003-2317-5471}\inst{\ref{aff63},\ref{aff64}}
\and I.~Lloro\orcid{0000-0001-5966-1434}\inst{\ref{aff67}}
\and M.~Magliocchetti\orcid{0000-0001-9158-4838}\inst{\ref{aff44}}
\and G.~Mainetti\orcid{0000-0003-2384-2377}\inst{\ref{aff68}}
\and O.~Mansutti\orcid{0000-0001-5758-4658}\inst{\ref{aff16}}
\and O.~Marggraf\orcid{0000-0001-7242-3852}\inst{\ref{aff69}}
\and M.~Martinelli\orcid{0000-0002-6943-7732}\inst{\ref{aff28},\ref{aff29}}
\and N.~Martinet\orcid{0000-0003-2786-7790}\inst{\ref{aff70}}
\and F.~Marulli\orcid{0000-0002-8850-0303}\inst{\ref{aff71},\ref{aff2},\ref{aff3}}
\and R.~J.~Massey\orcid{0000-0002-6085-3780}\inst{\ref{aff72}}
\and E.~Medinaceli\orcid{0000-0002-4040-7783}\inst{\ref{aff2}}
\and M.~Meneghetti\orcid{0000-0003-1225-7084}\inst{\ref{aff2},\ref{aff3}}
\and E.~Merlin\orcid{0000-0001-6870-8900}\inst{\ref{aff9}}
\and G.~Meylan\orcid{0000-0001-6503-0209}\inst{\ref{aff73}}
\and A.~Mora\orcid{0000-0002-1922-8529}\inst{\ref{aff74}}
\and M.~Moresco\orcid{0000-0002-7616-7136}\inst{\ref{aff71},\ref{aff2}}
\and C.~Moretti\orcid{0000-0003-3314-8936}\inst{\ref{aff16},\ref{aff15},\ref{aff17}}
\and L.~Moscardini\orcid{0000-0002-3473-6716}\inst{\ref{aff71},\ref{aff2},\ref{aff3}}
\and E.~Munari\orcid{0000-0002-1751-5946}\inst{\ref{aff16},\ref{aff15}}
\and R.~Nakajima\orcid{0009-0009-1213-7040}\inst{\ref{aff69}}
\and C.~Neissner\orcid{0000-0001-8524-4968}\inst{\ref{aff75},\ref{aff31}}
\and R.~C.~Nichol\orcid{0000-0003-0939-6518}\inst{\ref{aff76}}
\and S.-M.~Niemi\orcid{0009-0005-0247-0086}\inst{\ref{aff77}}
\and C.~Padilla\orcid{0000-0001-7951-0166}\inst{\ref{aff75}}
\and S.~Paltani\orcid{0000-0002-8108-9179}\inst{\ref{aff42}}
\and F.~Pasian\orcid{0000-0002-4869-3227}\inst{\ref{aff16}}
\and W.~J.~Percival\orcid{0000-0002-0644-5727}\inst{\ref{aff78},\ref{aff79},\ref{aff80}}
\and V.~Pettorino\orcid{0000-0002-4203-9320}\inst{\ref{aff77}}
\and A.~Pezzotta\orcid{0000-0003-0726-2268}\inst{\ref{aff14}}
\and S.~Pires\orcid{0000-0002-0249-2104}\inst{\ref{aff8}}
\and G.~Polenta\orcid{0000-0003-4067-9196}\inst{\ref{aff81}}
\and M.~Poncet\inst{\ref{aff82}}
\and L.~A.~Popa\inst{\ref{aff83}}
\and F.~Raison\orcid{0000-0002-7819-6918}\inst{\ref{aff51}}
\and A.~Renzi\orcid{0000-0001-9856-1970}\inst{\ref{aff84},\ref{aff85},\ref{aff2}}
\and J.~Rhodes\orcid{0000-0002-4485-8549}\inst{\ref{aff54}}
\and G.~Riccio\inst{\ref{aff23}}
\and E.~Romelli\orcid{0000-0003-3069-9222}\inst{\ref{aff16}}
\and M.~Roncarelli\orcid{0000-0001-9587-7822}\inst{\ref{aff2}}
\and C.~Rosset\orcid{0000-0003-0286-2192}\inst{\ref{aff86}}
\and B.~Rusholme\orcid{0000-0001-7648-4142}\inst{\ref{aff53}}
\and R.~Saglia\orcid{0000-0003-0378-7032}\inst{\ref{aff52},\ref{aff51}}
\and Z.~Sakr\orcid{0000-0002-4823-3757}\inst{\ref{aff87},\ref{aff88},\ref{aff89}}
\and A.~G.~S\'anchez\orcid{0000-0003-1198-831X}\inst{\ref{aff51}}
\and D.~Sapone\orcid{0000-0001-7089-4503}\inst{\ref{aff90}}
\and M.~Schirmer\orcid{0000-0003-2568-9994}\inst{\ref{aff59}}
\and P.~Schneider\orcid{0000-0001-8561-2679}\inst{\ref{aff69}}
\and A.~Secroun\orcid{0000-0003-0505-3710}\inst{\ref{aff61}}
\and E.~Sihvola\orcid{0000-0003-1804-7715}\inst{\ref{aff91}}
\and C.~Sirignano\orcid{0000-0002-0995-7146}\inst{\ref{aff84},\ref{aff85}}
\and G.~Sirri\orcid{0000-0003-2626-2853}\inst{\ref{aff3}}
\and L.~Stanco\orcid{0000-0002-9706-5104}\inst{\ref{aff85}}
\and P.~Tallada-Cresp\'{i}\orcid{0000-0002-1336-8328}\inst{\ref{aff30},\ref{aff31}}
\and I.~Tereno\orcid{0000-0002-4537-6218}\inst{\ref{aff92},\ref{aff93}}
\and S.~Toft\orcid{0000-0003-3631-7176}\inst{\ref{aff94},\ref{aff95}}
\and R.~Toledo-Moreo\orcid{0000-0002-2997-4859}\inst{\ref{aff96},\ref{aff97}}
\and F.~Torradeflot\orcid{0000-0003-1160-1517}\inst{\ref{aff31},\ref{aff30}}
\and I.~Tutusaus\orcid{0000-0002-3199-0399}\inst{\ref{aff47},\ref{aff46},\ref{aff88}}
\and E.~A.~Valentijn\inst{\ref{aff98}}
\and J.~Valiviita\orcid{0000-0001-6225-3693}\inst{\ref{aff63},\ref{aff64}}
\and T.~Vassallo\orcid{0000-0001-6512-6358}\inst{\ref{aff16},\ref{aff50}}
\and Y.~Wang\orcid{0000-0002-4749-2984}\inst{\ref{aff53}}
\and J.~Weller\orcid{0000-0002-8282-2010}\inst{\ref{aff52},\ref{aff51}}
\and A.~Zacchei\orcid{0000-0003-0396-1192}\inst{\ref{aff16},\ref{aff15}}
\and G.~Zamorani\orcid{0000-0002-2318-301X}\inst{\ref{aff2}}
\and F.~M.~Zerbi\orcid{0000-0002-9996-973X}\inst{\ref{aff14}}
\and E.~Zucca\orcid{0000-0002-5845-8132}\inst{\ref{aff2}}
\and T.~Castro\orcid{0000-0002-6292-3228}\inst{\ref{aff16},\ref{aff17},\ref{aff15},\ref{aff99}}}
										   
\institute{Institute for Astronomy, University of Edinburgh, Royal Observatory, Blackford Hill, Edinburgh EH9 3HJ, UK\label{aff1}
\and
INAF-Osservatorio di Astrofisica e Scienza dello Spazio di Bologna, Via Piero Gobetti 93/3, 40129 Bologna, Italy\label{aff2}
\and
INFN-Sezione di Bologna, Viale Berti Pichat 6/2, 40127 Bologna, Italy\label{aff3}
\and
Kobayashi-Maskawa Institute for the Origin of Particles and the Universe, Nagoya University, Chikusa-ku, Nagoya, 464-8602, Japan\label{aff4}
\and
Institute for Advanced Research, Nagoya University, Chikusa-ku, Nagoya, 464-8601, Japan\label{aff5}
\and
Kavli Institute for the Physics and Mathematics of the Universe (WPI), University of Tokyo, Kashiwa, Chiba 277-8583, Japan\label{aff6}
\and
Universit\'e Paris Cit\'e, Universit\'e Paris-Saclay, CEA, CNRS, AIM, F-91191, Gif-sur-Yvette, France\label{aff7}
\and
Universit\'e Paris-Saclay, Universit\'e Paris Cit\'e, CEA, CNRS, AIM, 91191, Gif-sur-Yvette, France\label{aff8}
\and
INAF-Osservatorio Astronomico di Padova, Via dell'Osservatorio 5, 35122 Padova, Italy\label{aff9}
\and
Universit\"at Innsbruck, Institut f\"ur Astro- und Teilchenphysik, Technikerstr. 25/8, 6020 Innsbruck, Austria\label{aff10}
\and
ESAC/ESA, Camino Bajo del Castillo, s/n., Urb. Villafranca del Castillo, 28692 Villanueva de la Ca\~nada, Madrid, Spain\label{aff11}
\and
Institut f\"ur Theoretische Physik, University of Heidelberg, Philosophenweg 16, 69120 Heidelberg, Germany\label{aff12}
\and
New York University Abu Dhabi, PO Box 129188, Abu Dhabi, UAE, and Center for Astrophysics and Space Science (CASS), New York University Abu Dhabi, UAE\label{aff13}
\and
INAF-Osservatorio Astronomico di Brera, Via Brera 28, 20122 Milano, Italy\label{aff14}
\and
IFPU, Institute for Fundamental Physics of the Universe, via Beirut 2, 34151 Trieste, Italy\label{aff15}
\and
INAF-Osservatorio Astronomico di Trieste, Via G. B. Tiepolo 11, 34143 Trieste, Italy\label{aff16}
\and
INFN, Sezione di Trieste, Via Valerio 2, 34127 Trieste TS, Italy\label{aff17}
\and
SISSA, International School for Advanced Studies, Via Bonomea 265, 34136 Trieste TS, Italy\label{aff18}
\and
Dipartimento di Fisica e Astronomia, Universit\`a di Bologna, Via Gobetti 93/2, 40129 Bologna, Italy\label{aff19}
\and
Dipartimento di Fisica, Universit\`a di Genova, Via Dodecaneso 33, 16146, Genova, Italy\label{aff20}
\and
INFN-Sezione di Genova, Via Dodecaneso 33, 16146, Genova, Italy\label{aff21}
\and
Department of Physics "E. Pancini", University Federico II, Via Cinthia 6, 80126, Napoli, Italy\label{aff22}
\and
INAF-Osservatorio Astronomico di Capodimonte, Via Moiariello 16, 80131 Napoli, Italy\label{aff23}
\and
Dipartimento di Fisica, Universit\`a degli Studi di Torino, Via P. Giuria 1, 10125 Torino, Italy\label{aff24}
\and
INFN-Sezione di Torino, Via P. Giuria 1, 10125 Torino, Italy\label{aff25}
\and
INAF-Osservatorio Astrofisico di Torino, Via Osservatorio 20, 10025 Pino Torinese (TO), Italy\label{aff26}
\and
INAF-IASF Milano, Via Alfonso Corti 12, 20133 Milano, Italy\label{aff27}
\and
INAF-Osservatorio Astronomico di Roma, Via Frascati 33, 00078 Monteporzio Catone, Italy\label{aff28}
\and
INFN-Sezione di Roma, Piazzale Aldo Moro, 2 - c/o Dipartimento di Fisica, Edificio G. Marconi, 00185 Roma, Italy\label{aff29}
\and
Centro de Investigaciones Energ\'eticas, Medioambientales y Tecnol\'ogicas (CIEMAT), Avenida Complutense 40, 28040 Madrid, Spain\label{aff30}
\and
Port d'Informaci\'{o} Cient\'{i}fica, Campus UAB, C. Albareda s/n, 08193 Bellaterra (Barcelona), Spain\label{aff31}
\and
INFN section of Naples, Via Cinthia 6, 80126, Napoli, Italy\label{aff32}
\and
Dipartimento di Fisica e Astronomia "Augusto Righi" - Alma Mater Studiorum Universit\`a di Bologna, Viale Berti Pichat 6/2, 40127 Bologna, Italy\label{aff33}
\and
Instituto de Astrof\'{\i}sica de Canarias, E-38205 La Laguna, Tenerife, Spain\label{aff34}
\and
European Space Agency/ESRIN, Largo Galileo Galilei 1, 00044 Frascati, Roma, Italy\label{aff35}
\and
Universit\'e Claude Bernard Lyon 1, CNRS/IN2P3, IP2I Lyon, UMR 5822, Villeurbanne, F-69100, France\label{aff36}
\and
Institut de Ci\`{e}ncies del Cosmos (ICCUB), Universitat de Barcelona (IEEC-UB), Mart\'{i} i Franqu\`{e}s 1, 08028 Barcelona, Spain\label{aff37}
\and
Instituci\'o Catalana de Recerca i Estudis Avan\c{c}ats (ICREA), Passeig de Llu\'{\i}s Companys 23, 08010 Barcelona, Spain\label{aff38}
\and
Institut de Ciencies de l'Espai (IEEC-CSIC), Campus UAB, Carrer de Can Magrans, s/n Cerdanyola del Vall\'es, 08193 Barcelona, Spain\label{aff39}
\and
UCB Lyon 1, CNRS/IN2P3, IUF, IP2I Lyon, 4 rue Enrico Fermi, 69622 Villeurbanne, France\label{aff40}
\and
Mullard Space Science Laboratory, University College London, Holmbury St Mary, Dorking, Surrey RH5 6NT, UK\label{aff41}
\and
Department of Astronomy, University of Geneva, ch. d'Ecogia 16, 1290 Versoix, Switzerland\label{aff42}
\and
Universit\'e Paris-Saclay, CNRS, Institut d'astrophysique spatiale, 91405, Orsay, France\label{aff43}
\and
INAF-Istituto di Astrofisica e Planetologia Spaziali, via del Fosso del Cavaliere, 100, 00100 Roma, Italy\label{aff44}
\and
INFN-Bologna, Via Irnerio 46, 40126 Bologna, Italy\label{aff45}
\and
Institut d'Estudis Espacials de Catalunya (IEEC),  Edifici RDIT, Campus UPC, 08860 Castelldefels, Barcelona, Spain\label{aff46}
\and
Institute of Space Sciences (ICE, CSIC), Campus UAB, Carrer de Can Magrans, s/n, 08193 Barcelona, Spain\label{aff47}
\and
School of Physics, HH Wills Physics Laboratory, University of Bristol, Tyndall Avenue, Bristol, BS8 1TL, UK\label{aff48}
\and
Department of Physics, Oxford University, Keble Road, Oxford OX1 3RH, UK\label{aff49}
\and
University Observatory, LMU Faculty of Physics, Scheinerstr.~1, 81679 Munich, Germany\label{aff50}
\and
Max Planck Institute for Extraterrestrial Physics, Giessenbachstr. 1, 85748 Garching, Germany\label{aff51}
\and
Universit\"ats-Sternwarte M\"unchen, Fakult\"at f\"ur Physik, Ludwig-Maximilians-Universit\"at M\"unchen, Scheinerstr.~1, 81679 M\"unchen, Germany\label{aff52}
\and
Caltech/IPAC, 1200 E. California Blvd., Pasadena, CA 91125, USA\label{aff53}
\and
Jet Propulsion Laboratory, California Institute of Technology, 4800 Oak Grove Drive, Pasadena, CA, 91109, USA\label{aff54}
\and
Department of Physics, Lancaster University, Lancaster, LA1 4YB, UK\label{aff55}
\and
Felix Hormuth Engineering, Goethestr. 17, 69181 Leimen, Germany\label{aff56}
\and
Technical University of Denmark, Elektrovej 327, 2800 Kgs. Lyngby, Denmark\label{aff57}
\and
Cosmic Dawn Center (DAWN), Denmark\label{aff58}
\and
Max-Planck-Institut f\"ur Astronomie, K\"onigstuhl 17, 69117 Heidelberg, Germany\label{aff59}
\and
NASA Goddard Space Flight Center, Greenbelt, MD 20771, USA\label{aff60}
\and
Aix-Marseille Universit\'e, CNRS/IN2P3, CPPM, Marseille, France\label{aff61}
\and
Universit\'e de Gen\`eve, D\'epartement de Physique Th\'eorique and Centre for Astroparticle Physics, 24 quai Ernest-Ansermet, CH-1211 Gen\`eve 4, Switzerland\label{aff62}
\and
Department of Physics, P.O. Box 64, University of Helsinki, 00014 Helsinki, Finland\label{aff63}
\and
Helsinki Institute of Physics, Gustaf H{\"a}llstr{\"o}min katu 2, University of Helsinki, 00014 Helsinki, Finland\label{aff64}
\and
Laboratoire d'etude de l'Univers et des phenomenes eXtremes, Observatoire de Paris, Universit\'e PSL, Sorbonne Universit\'e, CNRS, 92190 Meudon, France\label{aff65}
\and
Institute of Theoretical Astrophysics, University of Oslo, P.O. Box 1029 Blindern, 0315 Oslo, Norway\label{aff66}
\and
SKAO, Jodrell Bank, Lower Withington, Macclesfield SK11 9FT, UK\label{aff67}
\and
Centre de Calcul de l'IN2P3/CNRS, 21 avenue Pierre de Coubertin 69627 Villeurbanne Cedex, France\label{aff68}
\and
Universit\"at Bonn, Argelander-Institut f\"ur Astronomie, Auf dem H\"ugel 71, 53121 Bonn, Germany\label{aff69}
\and
Aix-Marseille Universit\'e, CNRS, CNES, LAM, Marseille, France\label{aff70}
\and
Dipartimento di Fisica e Astronomia "Augusto Righi" - Alma Mater Studiorum Universit\`a di Bologna, via Piero Gobetti 93/2, 40129 Bologna, Italy\label{aff71}
\and
Department of Physics, Institute for Computational Cosmology, Durham University, South Road, Durham, DH1 3LE, UK\label{aff72}
\and
Institute of Physics, Laboratory of Astrophysics, Ecole Polytechnique F\'ed\'erale de Lausanne (EPFL), Observatoire de Sauverny, 1290 Versoix, Switzerland\label{aff73}
\and
Telespazio UK S.L. for European Space Agency (ESA), Camino bajo del Castillo, s/n, Urbanizacion Villafranca del Castillo, Villanueva de la Ca\~nada, 28692 Madrid, Spain\label{aff74}
\and
Institut de F\'{i}sica d'Altes Energies (IFAE), The Barcelona Institute of Science and Technology, Campus UAB, 08193 Bellaterra (Barcelona), Spain\label{aff75}
\and
School of Mathematics and Physics, University of Surrey, Guildford, Surrey, GU2 7XH, UK\label{aff76}
\and
European Space Agency/ESTEC, Keplerlaan 1, 2201 AZ Noordwijk, The Netherlands\label{aff77}
\and
Waterloo Centre for Astrophysics, University of Waterloo, Waterloo, Ontario N2L 3G1, Canada\label{aff78}
\and
Department of Physics and Astronomy, University of Waterloo, Waterloo, Ontario N2L 3G1, Canada\label{aff79}
\and
Perimeter Institute for Theoretical Physics, Waterloo, Ontario N2L 2Y5, Canada\label{aff80}
\and
Space Science Data Center, Italian Space Agency, via del Politecnico snc, 00133 Roma, Italy\label{aff81}
\and
Centre National d'Etudes Spatiales -- Centre spatial de Toulouse, 18 avenue Edouard Belin, 31401 Toulouse Cedex 9, France\label{aff82}
\and
Institute of Space Science, Str. Atomistilor, nr. 409 M\u{a}gurele, Ilfov, 077125, Romania\label{aff83}
\and
Dipartimento di Fisica e Astronomia "G. Galilei", Universit\`a di Padova, Via Marzolo 8, 35131 Padova, Italy\label{aff84}
\and
INFN-Padova, Via Marzolo 8, 35131 Padova, Italy\label{aff85}
\and
Universit\'e Paris Cit\'e, CNRS, Astroparticule et Cosmologie, 75013 Paris, France\label{aff86}
\and
Instituto de F\'isica Te\'orica UAM-CSIC, Campus de Cantoblanco, 28049 Madrid, Spain\label{aff87}
\and
Institut de Recherche en Astrophysique et Plan\'etologie (IRAP), Universit\'e de Toulouse, CNRS, UPS, CNES, 14 Av. Edouard Belin, 31400 Toulouse, France\label{aff88}
\and
Universit\'e St Joseph; Faculty of Sciences, Beirut, Lebanon\label{aff89}
\and
Departamento de F\'isica, FCFM, Universidad de Chile, Blanco Encalada 2008, Santiago, Chile\label{aff90}
\and
Department of Physics and Helsinki Institute of Physics, Gustaf H\"allstr\"omin katu 2, University of Helsinki, 00014 Helsinki, Finland\label{aff91}
\and
Departamento de F\'isica, Faculdade de Ci\^encias, Universidade de Lisboa, Edif\'icio C8, Campo Grande, PT1749-016 Lisboa, Portugal\label{aff92}
\and
Instituto de Astrof\'isica e Ci\^encias do Espa\c{c}o, Faculdade de Ci\^encias, Universidade de Lisboa, Tapada da Ajuda, 1349-018 Lisboa, Portugal\label{aff93}
\and
Cosmic Dawn Center (DAWN)\label{aff94}
\and
Niels Bohr Institute, University of Copenhagen, Jagtvej 128, 2200 Copenhagen, Denmark\label{aff95}
\and
Universidad Polit\'ecnica de Cartagena, Departamento de Electr\'onica y Tecnolog\'ia de Computadoras,  Plaza del Hospital 1, 30202 Cartagena, Spain\label{aff96}
\and
European University of Technology EUt+, European Union\label{aff97}
\and
Kapteyn Astronomical Institute, University of Groningen, PO Box 800, 9700 AV Groningen, The Netherlands\label{aff98}
\and
ICSC - Centro Nazionale di Ricerca in High Performance Computing, Big Data e Quantum Computing, Via Magnanelli 2, Bologna, Italy\label{aff99}}     

%
%
\abstract{
We present a \lensmc lensing analysis of \Euclid Quick Release 1 images that were made available in March 2025.
We measured shapes, positions, weights, and other morphological parameters of galaxies with a surface number density of $26\;\text{arcmin}^{-2}$ for $\IE<24.5$, achieving $75\;\unit{arcmin^{-2}}$ for $\IE<27$, in 63\;deg$^2$ of \Euclid VIS images.
This is the first shear measurement catalogue produced with \lensmc in anticipation of the Euclid Data Release 1.
To within the scope of this work and availability of survey volume, we found no spatial dependency in the additive biases.
However, we developed an empirical bias correction based jointly on galaxy sizes and magnitudes, which was applied to the measurement of the lensing signal over the whole area.
In order to validate the quality of our measurements, we calculated two-point statistics and cluster profile measurements, and cross-checked results with external cluster catalogues from WISE and the Dark Energy Survey.
Additionally, by stacking shear profiles of random clusters we found that the low-redshift, large radii bins may be still contaminated by residual systematic effects.
Thanks to \Euclid image resolution and depth and overall good control of systematic errors, we are able to constrain the lensing profiles of clusters with masses of $10^{14}M_\odot$ out to $z\approx2$ over nearly 10\;Gyr of evolution history.
}
%
%
\keywords{
    Gravitational lensing: weak --
    Galaxies: clusters: general --
    Cosmology: observations --
    Methods: data analysis
    }
%
%
   \titlerunning{LensMC catalogue for Euclid Q1 cluster lensing}
   \authorrunning{G.~Congedo et al.}
   
\maketitle
%
%
%
%
   
\section{Introduction}\label{sec:introduction}

\Euclid is poised to release 1900\;deg$^2$ of Data Release 1 (DR1) images by the end of 2026 \citep{Laureijs11,Scaramella-EP1,EuclidSkyOverview}.
The optical instrument \citep[VIS,][]{EuclidSkyVIS} and near-infrared instrument \citep[NISP,][]{EuclidSkyNISP} are delivering exquisite images that were showcased in the \Euclid Early Release Observations \citep[ERO, 10\;deg$^2$,][]{EROData,EROcite} and Euclid Quick Release 1 \citep[Q1, 63\;deg$^2$,][]{Q1-TP001,Q1cite}.

In addition to the percent-level constraints on cosmological models that \Euclid will provide, thanks to weak lensing and galaxy clustering \citep[see relevant sections in][]{EuclidSkyOverview}, \Euclid is already revolutionising astronomy with its rich science. Particularly interesting is the case of galaxy clusters.
Forming at the nodes of the cosmic web, their lensing profiles and mass function can be used to constrain $\Omm$ and $\sigma_8$, independently from cosmic shear \citep[see the extensive reviews by][]{mandelbaum2018,umetsu2020,miyatake2025}.

As part of the successful ERO campaign, \Euclid observed two massive clusters: Perseus \citep{EROPerseusOverview,EROPerseusDGs,EROPerseusICL}; and Abell 2390 \citep{EROLensData}.
Regarding Abell 2390, an extensive lensing analysis was presented for 0.57\;deg$^2$ of deep data with three shape measurement methods, accurate redshift calibration, and cluster member decontamination \citep{Schrabback25}.
Furthermore, combining weak and strong lensing allows us to access both the inner and outer regions of the cluster \citep{Diego25}. 
With the inclusion of additional spectroscopic data, results can be further refined \citep{abriola2025}.
Finally, \citet{Ellien25} showed the complementarity of combining information from intracluster light with lensing and X-rays data.

Cluster cosmology is an integral part of the rich science that \Euclid is delivering.
Recent work includes optimally detecting clusters for cosmology with number counts \citep{Adam-EP3},
modelling the uncertainties in counts \citep{Fumagalli21} and two-point correlation functions of cluster positions \citep{Fumagalli-EP27},
estimating cluster masses with external data \citep{Giocoli-EP30},
accounting for the correct selection of background galaxies via colour cuts \citep{EP-Lesci},
assessing the impact of line-of-sight projections with multi-band data \citep{EP-Ragagnin},
using Q1 data to understand the relation between clusters and filaments \citep{Q1-SP005},
and compiling the first catalogue of strongly-lensing clusters \citep{Q1-SP057}.
A reanalysis of Stage-III data with \Euclid methods showed that \Euclid is expected to detect of order 10\,000 clusters with good control of systematic errors \citep{EP-Sereno}.
However, cluster masses inferred from \Euclid-like simulations can still be  biased by -15\% \citep{EP-Ingoglia}.
Provided that cluster observable systematic effects are kept under control, the clustering with number counts of clusters can also be used as additional cosmological probes, hence breaking degeneracies in cosmological parameters, especially when combined \citep{fumagalli2025}.
It will be essential to validate the \Euclid cluster catalogue with external catalogues \citep{EP-Melin}.
The observation of galaxy over-densities in Q1 data allowed the detection of a number of clusters in \Euclid images, some of which were completely new and some were cross-validated with X-rays and Sunyaev–Zeldovich observations \citep{Q1-SP050}.
However, the quality of such catalogues will be reassessed with DR1.

\begin{figure*}
\centering
\begin{overpic}[width=0.8\textwidth]{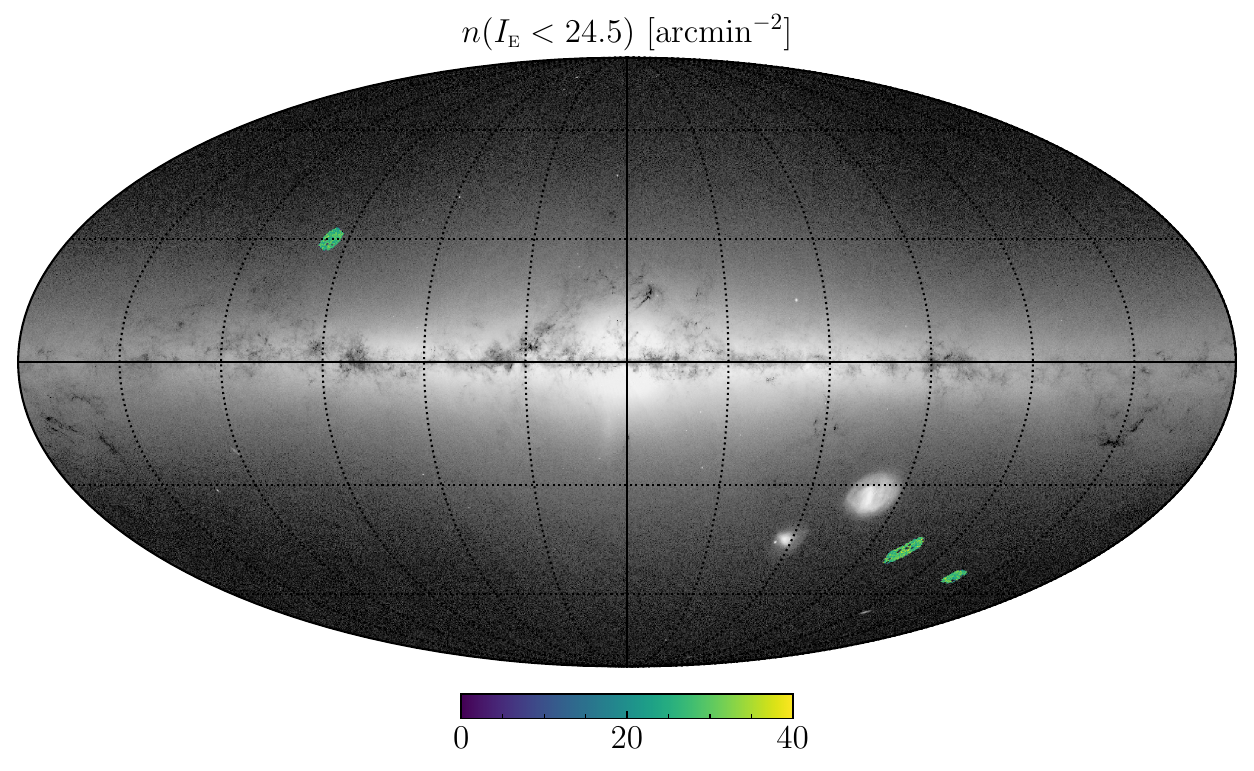}
\put(-10,40){\fcolorbox{black}{white}{\begin{tabular}{c}
EDF-North \\
\includegraphics[trim=30 70 0 30,clip,width=0.2\linewidth]{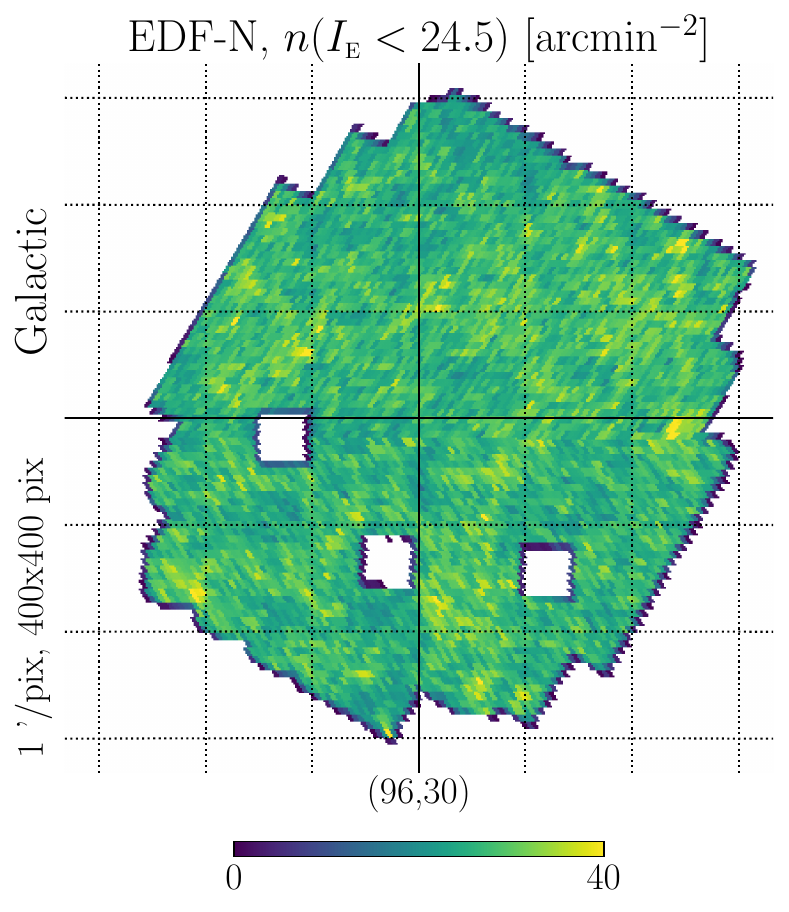}
\end{tabular}}}
\put(75,48){\fcolorbox{black}{white}{\begin{tabular}{c}
EDF-South \\
\includegraphics[trim=30 70 0 30,clip,width=0.2\linewidth]{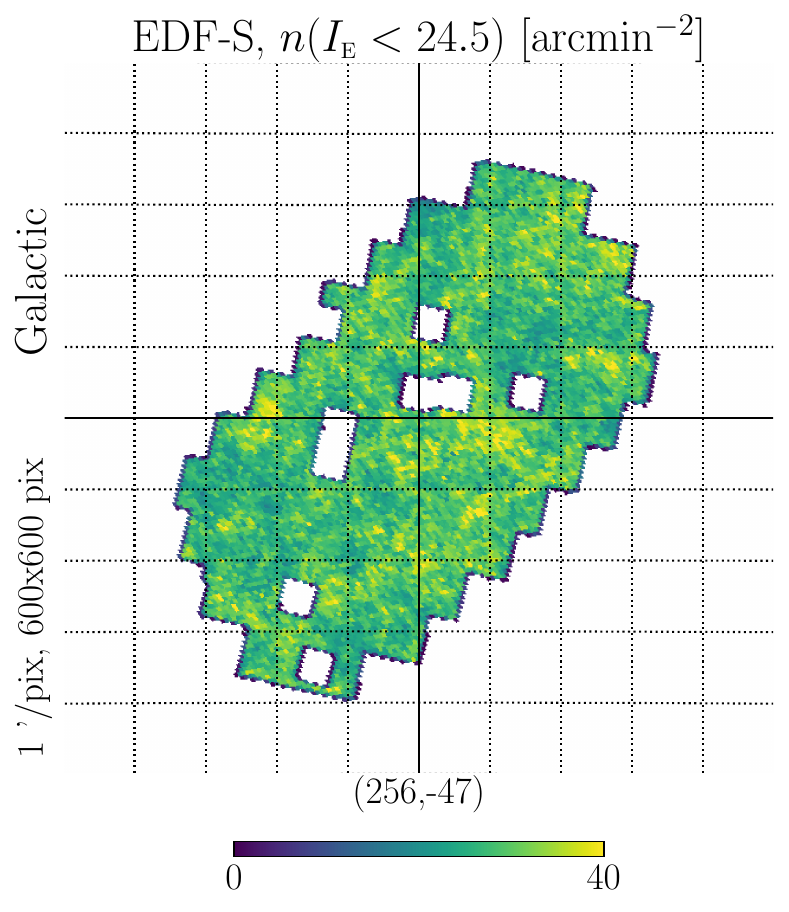}
\end{tabular}}}
\put(80,16){\fcolorbox{black}{white}{\begin{tabular}{c}
EDF-Fornax \\
\includegraphics[trim=30 70 0 30,clip,width=0.2\linewidth]{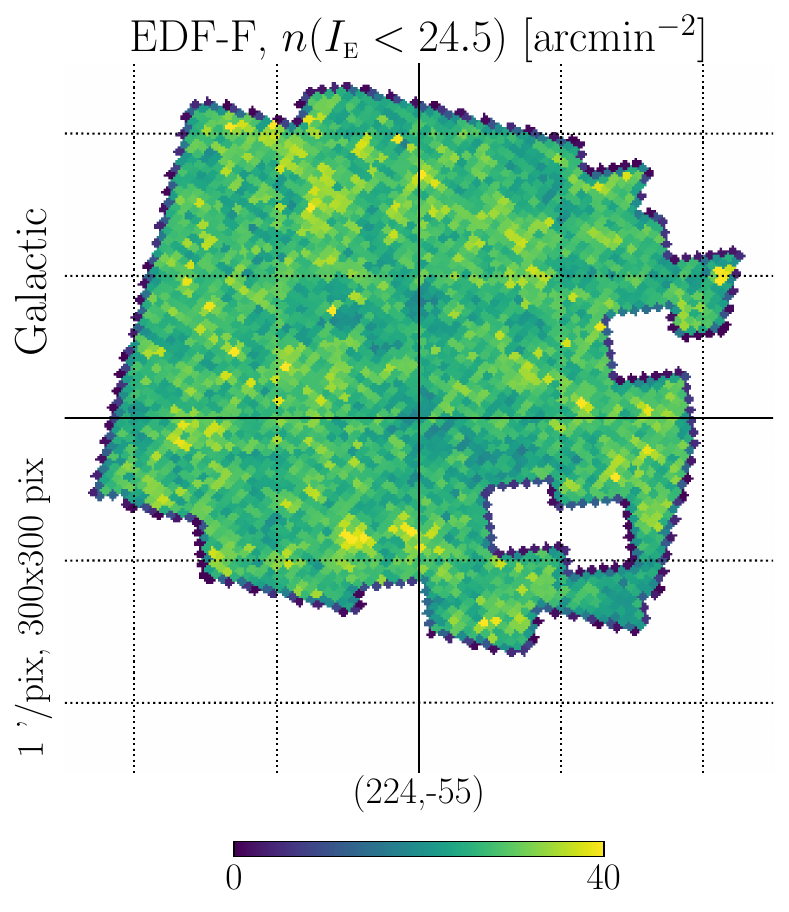}
\end{tabular}}}
\end{overpic}
\caption{Number density of galaxies processed by \lensmc after the quality cuts.
The image insets show the three Euclid Q1 fields: North ($\alpha=\ang{270}$, $\delta=\ang{66}$), South ($\alpha=\ang{61}$, $\delta=\ang{-48}$), and Fornax ($\alpha=\ang{53}$, $\delta=\ang{-28}$),
with areas of 23\;deg$^2$, 28\;deg$^2$, and 12\;deg$^2$,
totalling 63\;deg$^2$.
The mean number density is $26\;\text{arcmin}^{-2}$ ($\IE<24.5$) after the input and post-measurement quality cuts.
Star counts are from \emph{Gaia} DR3-data \citep{vallenari2023}.}
\label{fig:footprint}
\end{figure*}

This paper presents a shear measurement catalogue of the three Q1 fields for lensing applications.
We employed the official cosmic shear method for Euclid DR1 \citep[\lensmc,][hereafter \citetalias{EP-Congedo}]{EP-Congedo}.
After the successful application of \lensmc to a relatively small area of the ERO data set, this is the first analysis of a more substantial area of 63\;deg$^2$ (see Fig.\;\ref{fig:footprint}), ahead of the anticipated DR1 area.
Thanks to the quality of the data and good control of systematic errors, we demonstrate our ability to constrain, for the first time, the lensing profiles of clusters with a relatively low mass of $10^{14}M_\odot$ out to $z\approx2$ over nearly 10\;Gyr of evolution history.
Section\;\ref{sec:measurement} describes the measurements and production and validation of the catalogue.
Section\;\ref{sec:bias} presents tests on additive biases and correction of the catalogue.
Section\;\ref{sec:profiles} summarises the main validation tests that we carried out on external cluster catalogues that we observed in Q1.
Section\;\ref{sec:conclusions} draws the main conclusions of our work.

\section{Shear measurements}\label{sec:measurement}

\lensmc was specifically designed to meet the cosmic shear requirements for \Euclid and Stage-IV galaxy surveys,
and tested against pre-launch simulations \citepalias{EP-Congedo}.
These requirements determine the maximum acceptable statistical uncertainty on residual multiplicative bias, $m$, and additive bias, $c$ (see the linear modelling of equations 8 and 9 in \citetalias{EP-Congedo}): $\sigma_m=2\times10^{-3}$ and $\sigma_c=3\times10^{-4}$, which were established during the mission adoption phase \citep{massey2013,cropper2013} and based on earlier results \citep{amara2008}.
The requirements represent the statistical uncertainty to which the raw, uncalibrated shear would need to be measured in order to be successfully calibrated.
Although the \Euclid requirements were determined only for the final, full DR3 survey area, which is nominally 14\,000\;deg$^2$,
benchmark figures can still be inferred based on expected footprint areas as reported in Table\;\ref{tab:requirements}.
Specifically, we calculated the following reference requirements: $\sigma_{m,\,\sfont{DR1}}=5.7\times10^{-3}$ and $\sigma_{c,\,\sfont{DR1}}=8.6\times10^{-4}$ for DR1 (conservatively assuming an area of 1700\;deg$^2$), and $\sigma_{m,\,\sfont{Q1}}=3.0\times10^{-2}$ and $\sigma_{c,\,\sfont{Q1}}=4.5\times10^{-3}$ for Q1.
These figures are relevant for weak lensing cosmic shear and, as such, are extremely conservative for applications to cluster lensing, which can tolerate residual systematic errrors of the order of $1\%$ or even greater \citep{kohlinger2015}.
In comparison, raw multiplicative biases for \lensmc were found to be 1\% or less, including when accounting for model biases arising from model incompleteness in the galaxy bulge component, in fiducial pre-launch simulations between input magnitudes of 20 and 25 \citepalias{EP-Congedo}.
In contrast, this level of bias is more than a factor 3 better than the Q1 requirement derived above, so we rest assured that our methodology is good enough for cluster lensing in the Q1 area.

\begin{table}[h]
\caption{\Euclid requirements on cosmic shear measurements for the three data releases and first quick release analysed here.
The derived cosmic shear requirement for Q1 are more than an order of magnitude larger than DR3, and deemed to be conservative for cluster lensing measurements.}
\label{tab:requirements}
\centering
\setlength{\tabcolsep}{3pt}
\begin{tabular}{c@{\hskip 10pt}c@{\hskip 10pt}c@{\hskip 10pt}c@{\hskip 10pt}c}
\hline\hline
& \text{Q1} & \text{DR1} & \text{DR2} & \text{DR3} \\
& \text{(63\;deg$^2$)} & \text{(1700\;deg$^2$)} & \text{(3700\;deg$^2$)} & \text{(14\,000\;deg$^2$)} \\
\hline
$\sigma_m$ & $3\times10^{-2}$ & $5.7\times10^{-3}$ & $3.9\times10^{-3}$ & $2\times10^{-3}$ \\
$\sigma_c$ & $4.5\times10^{-3}$ & $8.6\times10^{-4}$ & $5.8\times10^{-4}$ & $3\times10^{-4}$ \\
\hline\hline
\end{tabular}
\end{table}

The Q1 release consists of three Euclid Deep Fields (EDFs): EDF-South ($\alpha=\ang{61}$, $\delta=\ang{-48}$), EDF-Fornax ($\alpha=\ang{53}$, $\delta=\ang{-28}$), and EDF-North ($\alpha=\ang{270}$, $\delta=\ang{66}$).
We processed the following data products contained in the official release.
\begin{enumerate}
\item Mosaic stacked images `\texttt{EUC\_MER\_BGSUB-MOSAIC-VIS}', PSF images `\texttt{EUC\_MER\_GRID-PSF-VIS}', and pixel flag maps `\texttt{EUC\_MER\_MOSAIC-VIS-FLAG}' \citep[MER,][]{Q1-TP004} based on the reprocessing of the raw image frames \citep[VIS,][]{Q1-TP002}.
\item Detection catalogues `\texttt{EUC\_MER\_FINAL-CAT}' and segmentation maps `\texttt{EUC\_MER\_FINAL-SEGMAP}' \citep[MER,][]{Q1-TP004}.
\item Photometric redshifts `\texttt{EUC\_PHZ\_PHZCAT}' \citep[PHZ,][]{Q1-TP003,Q1-TP005}.
\end{enumerate}
We applied \lensmc to the stacked images, using the supplied PSF images, maps, and flags.
As for the main weak lensing analysis, this analysis was restricted to objects detected in VIS, rather than NISP.
Additionally, bad image pixels were masked with a VIS flag value of one.
The released PSF images consist of a regular grid with a coarseness of \ang{;;12} over the entire stacked image of \ang{;32} in size.
In order to generate an oversampled image of the PSF at an arbitrary position in the stacked image during the shear measurement, we employed the following strategy:
\begin{enumerate}
\item cut out postage stamps of size \ang{;;3.2} around the provided PSF positions in the grid;
\item identify the presence of flagged pixels within \ang{;;0.3} from the centre of the PSF image, enough to encompass the core of the PSF;
\item upsample the unflagged PSF cutouts by spline interpolation to an oversampling factor of 3 to avoid biases from undersampling in the galaxy model generation \citepalias{EP-Congedo};
\item generate the PSF image at any given position in the FoV by bilinear interpolation or, when not possible, by nearest-neighbour extrapolation.\footnote{In general, L\'anczos interpolation might be the preferred choice; however, at the expense of potentially introducing ringing.
Since the supplied PSF model is relatively smooth, bilinear interpolation is still enough for the purposes of this work.}
\end{enumerate}
With the PSF generation in place, we measured all the detected objects from 342 input catalogue tiles covering the three fields,
with the same methodology applied to pre-launch simulations \citepalias{EP-Congedo} and with the improvements that were made following the application to the ERO images \citep{Schrabback25,Diego25}.
Finally, we produced a catalogue resulting from the merging of all the tile catalogues, with the addition of a few columns from the MER and PHZ data products and further derived quantities.

\begin{figure}
\centering
\includegraphics[width=\columnwidth]{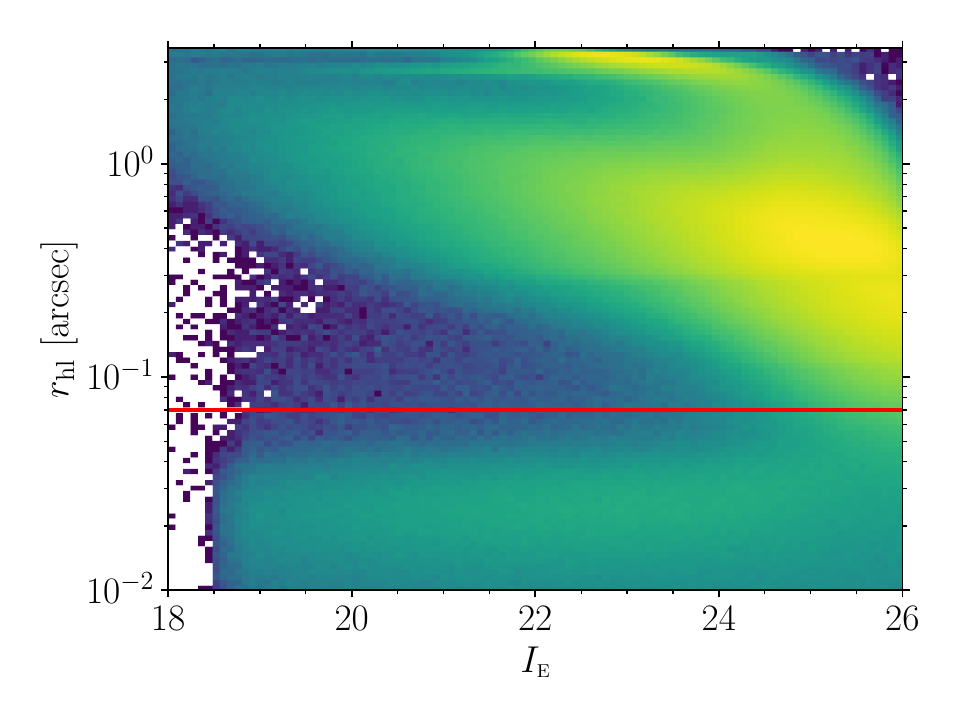}
\caption{Joint distribution of \lensmc magnitude and flux-averaged half-light radius.
The horizontal line denotes the main selection applied to the catalogue to remove stars after the quality cuts to the catalogue.
The secondary branch above $\ang{;;2}$ is due to spurious inference of the size parameter (see Sect.\;\ref{sec:measurement} for details).
Counts are reported on a logarithmic scale. }
\label{fig:mag-size}
\end{figure}

Figure\;\ref{fig:mag-size} shows the magnitude--size distribution of the \lensmc estimates after the quality selection cuts applied at the level of detection and shear measurement catalogues, hence excluding objects that are most likely associated with false detections.
Although all detected objects were measured by \lensmc (hence shapes and redshifts are still available in the catalogue), we applied a quality cut by retaining objects flagged with \texttt{DET\_QUALITY\_FLAG}=0, 1, 2, or 512, and \texttt{SPURIOUS\_FLAG}=0. 
This selection effectively reduces the catalogue to $84\%$ of its original size, excluding objects that are saturated, too close to the edge of the detector, within a bright star mask, failed the deblending stage, or were flagged as spurious according to a defined probability threshold trained on pre-launch simulations \citep{Q1-TP004}.
Figure\;\ref{fig:mag-size} also shows the star--galaxy separation that was applied to the measured \lensmc catalogue, consisting of a cut on the measured flux-averaged half-light radius.
This is consistent with our pre-launch investigations, where a selection based on the inferred object size was found to be optimal for star--galaxy separation, with extremely low levels of contamination from false positives and false negatives \citepalias{EP-Congedo}.
Additionally, the small-size branch associated with stars appears to be flat with magnitude, and the point of confusion where the two populations merge is at $\IE\approx25$.
This gives us great reassurance that the contamination is insignificant, again consistent with findings from our pre-launch simulations.
The magnitude-independent cut that we adopted is above $r_\text{hl}=\ang{;;0.07}$.
The secondary branch above $\approx\ang{;;2}$ is due to a combination of faint sources being incorrectly reported as having a large size and genuine bulge-dominated galaxies being incorrectly fitted in the modelling.
In fact, since \lensmc fixes the bulge-to-disc size ratio to a value informed by external measurements \citep{welikala2025}, this assumption was verified to lead to an additional model bias of $|m_\text{model}|<1\%$.
However, it was proved that this model bias could be calibrated due to its low sensitivity to the simulation setup \citepalias{EP-Congedo}.
For DR1, a more aggressive selection based on a magnitude-dependent cut will be applied, with any residual multiplicative biases estimated via image-simulation calibration. 
A further selection comes from removing a very small fraction of objects whose measurements failed due to the quality of the input data or inconsistencies in the provided maps.

\begin{figure}
\centering
\includegraphics[width=\columnwidth]{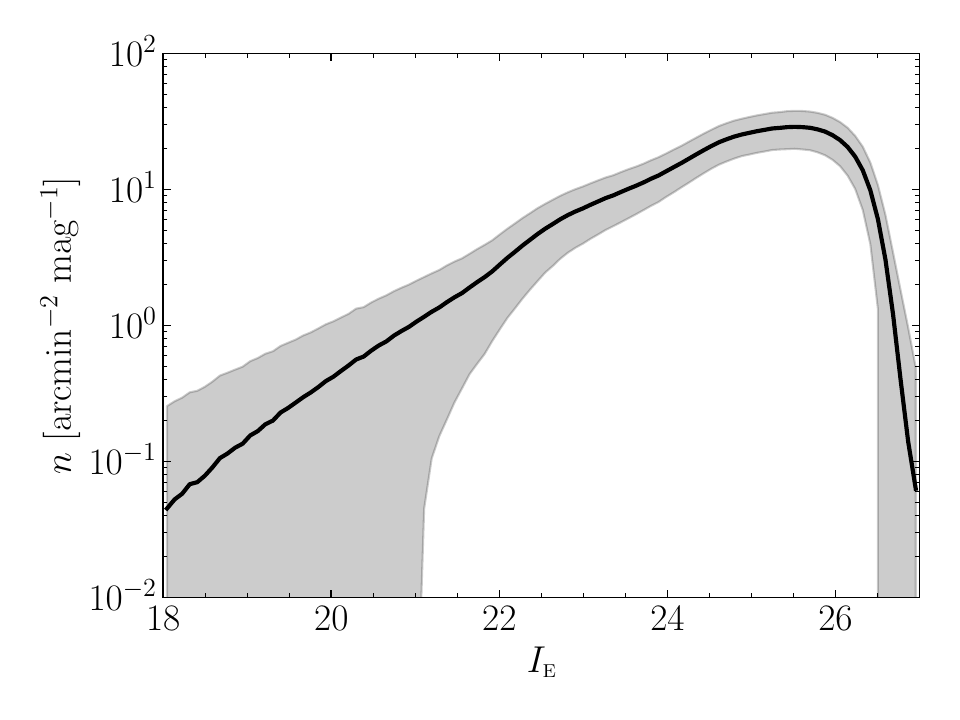} \\
\includegraphics[width=\columnwidth]{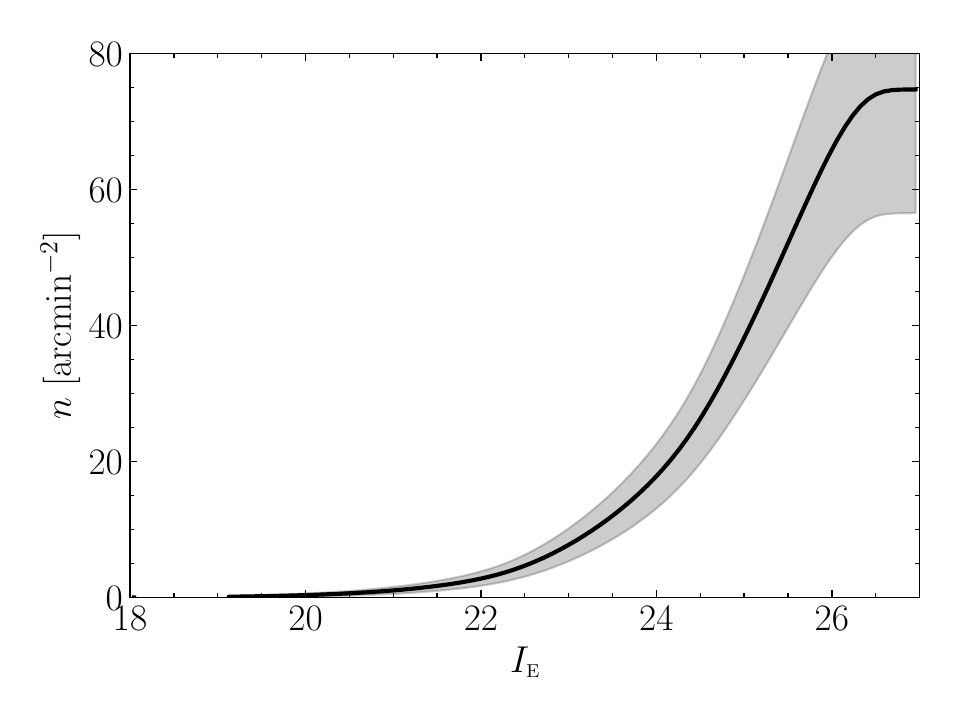}
\caption{Number counts of galaxies after the \lensmc quality cuts, differentially (\emph{top}) and cumulatively (\emph{bottom}).
The catalogue appears to be complete up to $\IE\approx25$.
The total mean count of galaxies is $\approx26\;\text{arcmin}^{-2}$ ($\IE<24.5$).
The statistical error (shown as the band around the mean) is the 1-$\sigma$ standard deviation across all the \healpix pixels in the Q1 fields.}
\label{fig:count}
\end{figure}

Figure\;\ref{fig:count} shows the number count of galaxies estimated in \healpix \citep{gorski2005, zonca2019} pixels of $11.8\;\text{arcmin}^2$ (\texttt{nside}=1024) along with its statistical variation across the Q1 fields.
The catalogue appears complete up to $\IE\approx25$, with a total number count of $(26.3\pm6.6)\;\text{arcmin}^{-2}$ for $\IE<24.5$, and achieving $\approx75\;\unit{arcmin^{-2}}$ for $\IE<27$ after all the quality cuts.
It is worth highlighting that both the shape of the magnitude count and total counts are in agreement with the pre-launch simulations.

\section{Bias correction}\label{sec:bias}

Motivated by stringent science requirements (see Table\;\ref{tab:requirements}), the PSF modelling in the official weak lensing analysis for DR1 is based on forward modelling the wavefront as it propagates through the telescope optical components.
This model also predicts the PSF images in every exposure, depending on the position in the field of view (FoV) and spectral energy distribution of the object (Miller et al., in prep.; Duncan et al., in prep.).
In contrast, because our analysis is bound to requirements that are primarily driven by the specific science goals of cluster lensing and the very limited data volume,
our Q1 requirements are more than an order of magnitude looser than for the weak lensing analysis. 
With that in mind, we still proceeded by testing the accuracy of our PSF modelling that is based on interpolating over the PSF image grids that are provided with the stacked images.

\begin{figure}
\centering
\includegraphics[width=\columnwidth]{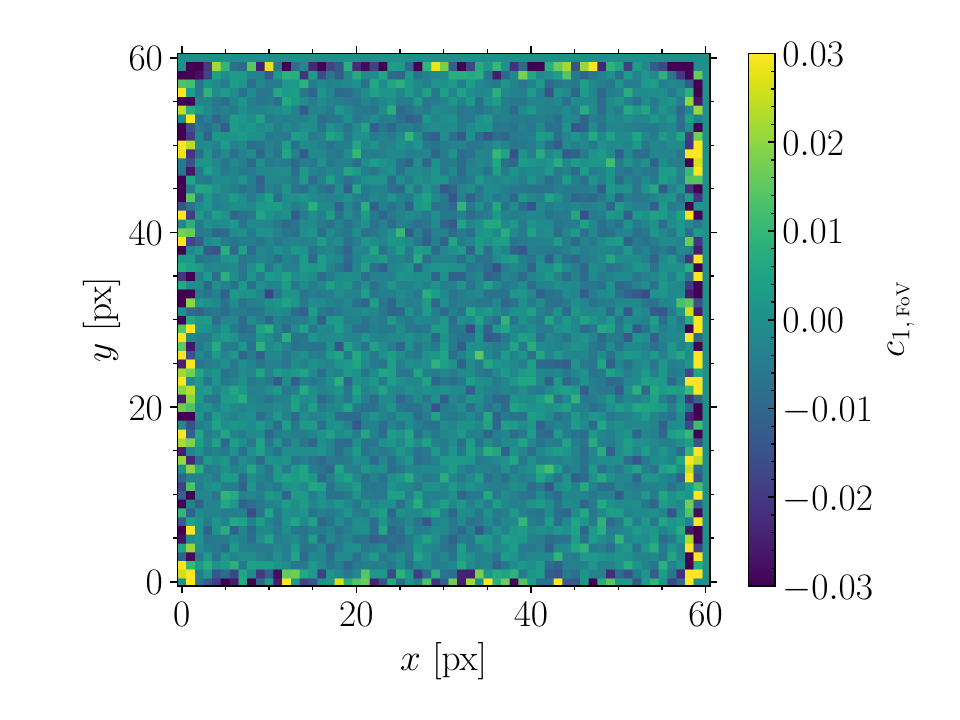} \\
\includegraphics[width=\columnwidth]{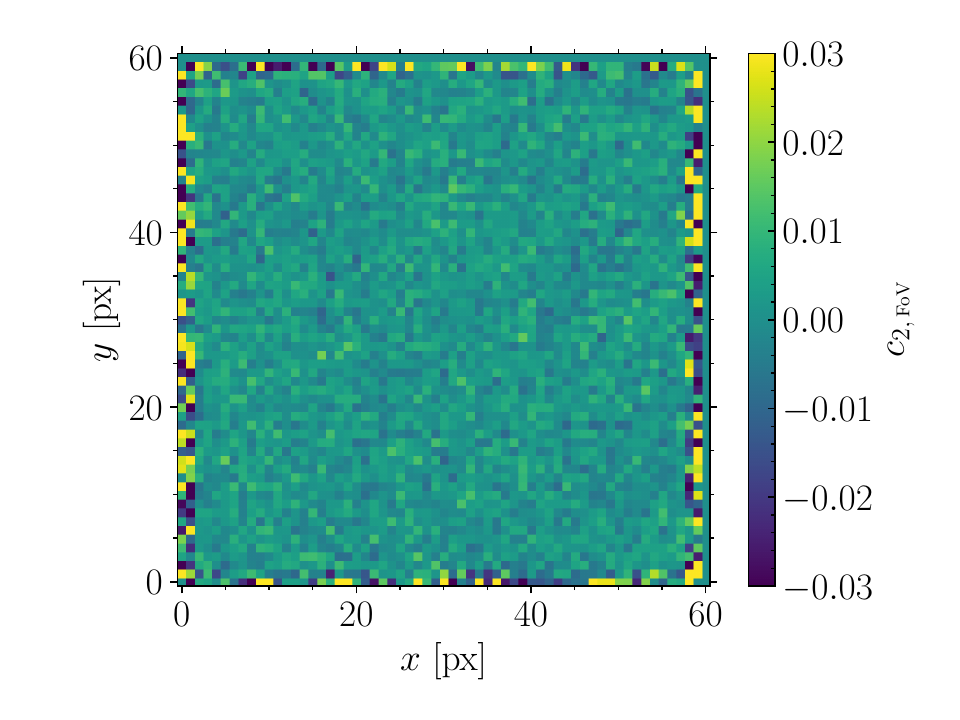} \\
\includegraphics[width=\columnwidth]{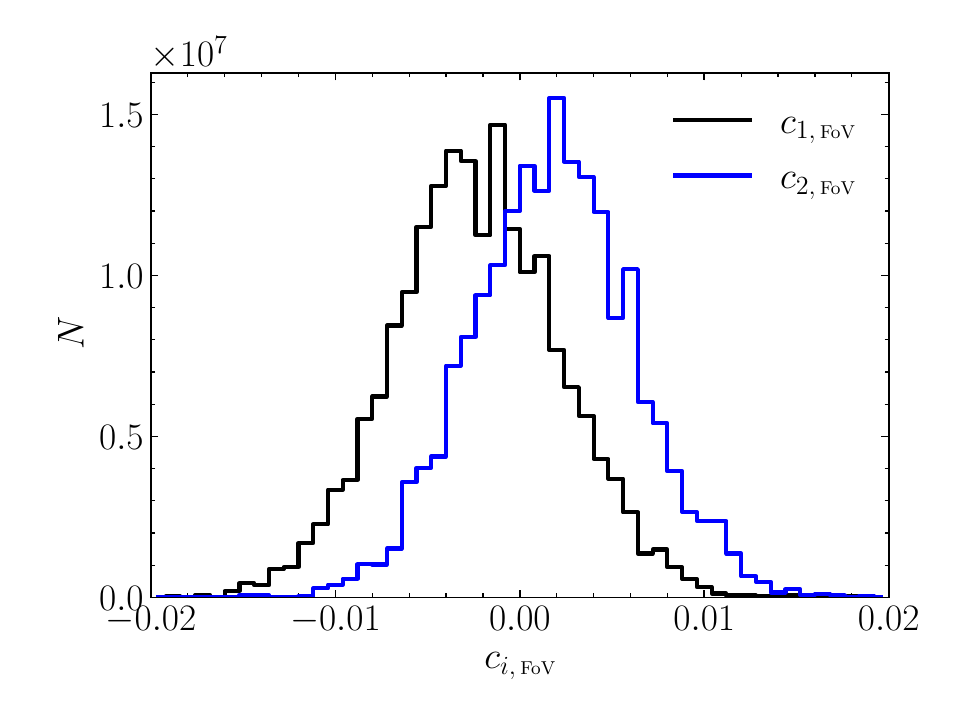}
\caption{Additive biases in field-of-view coordinates.
Although no spatial pattern is present in the maps (\emph{top and middle panels}), a clear shift can be observed in the 1D distributions of $c_{1,\,\sfont{FoV}}\approx-2\times10^{-3}$ and $c_{2,\,\sfont{FoV}}=2\times10^{-3}$ (\emph{bottom panel}).
The image scale is $\ang{;0.5}$ / pixel.}
\label{fig:c_fov_map}
\end{figure}

As a first step, we calculated a map of the shear components in the FoV plane.
By construction, the FoV plane is aligned with the world coordinate frame, since the stacking process consists of aligning and stacking all the exposures to the same coordinate frame.
Therefore, the reported \lensmc ellipticities will be conveniently defined in both world coordinate and FoV frames, without requiring any rotation between frames.
The measured positions were converted to FoV pixel positions using the supplied astrometric solution and all shapes were finally averaged in FoV position bins across the three Q1 fields.
The derived shear bias map, which does not contain cosmological signal to first order and at the scales of interest, is a direct measurement of $c$-biases.
We did not attempt to study the fields individually because the shear bias maps are in general noisy due to the small data volume.
Of course, the procedure does not capture $m$-biases, which can only be measured via simulations or via the recently proposed forward-modelling self-calibration method \citep[see extensive discussion in][]{congedo2025}.
The results of the FoV analysis are shown in Fig.\;\ref{fig:c_fov_map}.
Reassuringly, no apparent systematic pattern can be observed across the FoV, apart from the increased variance at the edge of the map due to having fewer objects in those position bins.
The apparent lack of spatial variability is due to a combination of small data volume and the image stacking procedure, which may be diluting the signal.
The distributions of $c$-biases are visibly shifted from zero, and the calculated values are: $c_{1,\,\sfont{FoV}}=(-2.35\pm0.08)\times10^{-3}$ and $c_{2,\,\sfont{FoV}}=(1.79\pm0.08)\times10^{-3}$.
The seeming absence of spatial variability should give us reassurance of the ability to calibrate the effect within the scope of this work.

\begin{figure}
\centering
\includegraphics[width=\columnwidth]{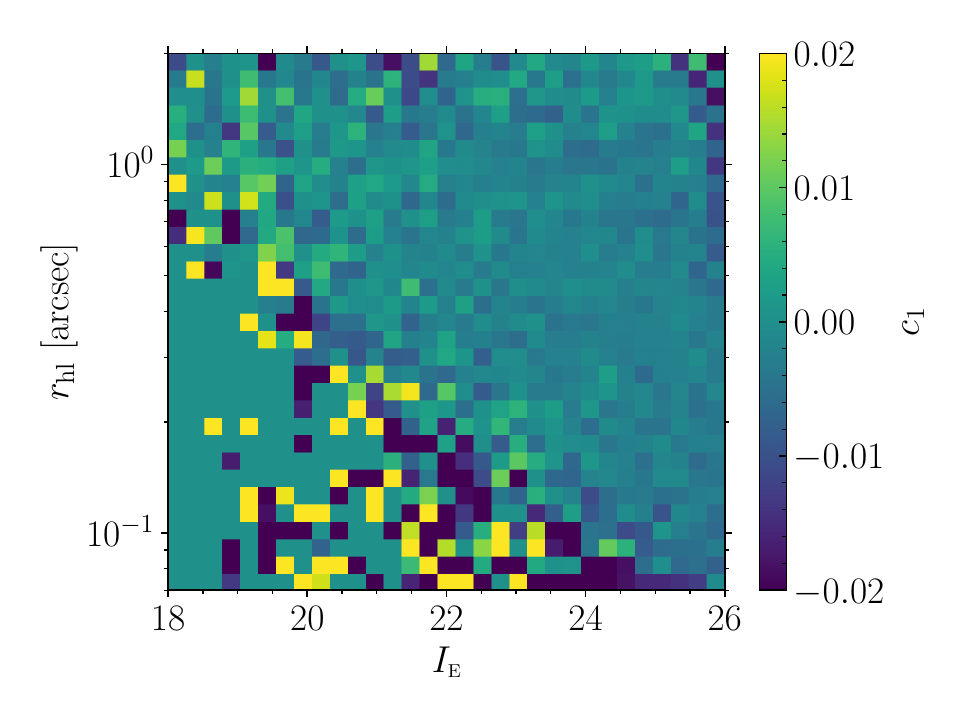} \\
\includegraphics[width=\columnwidth]{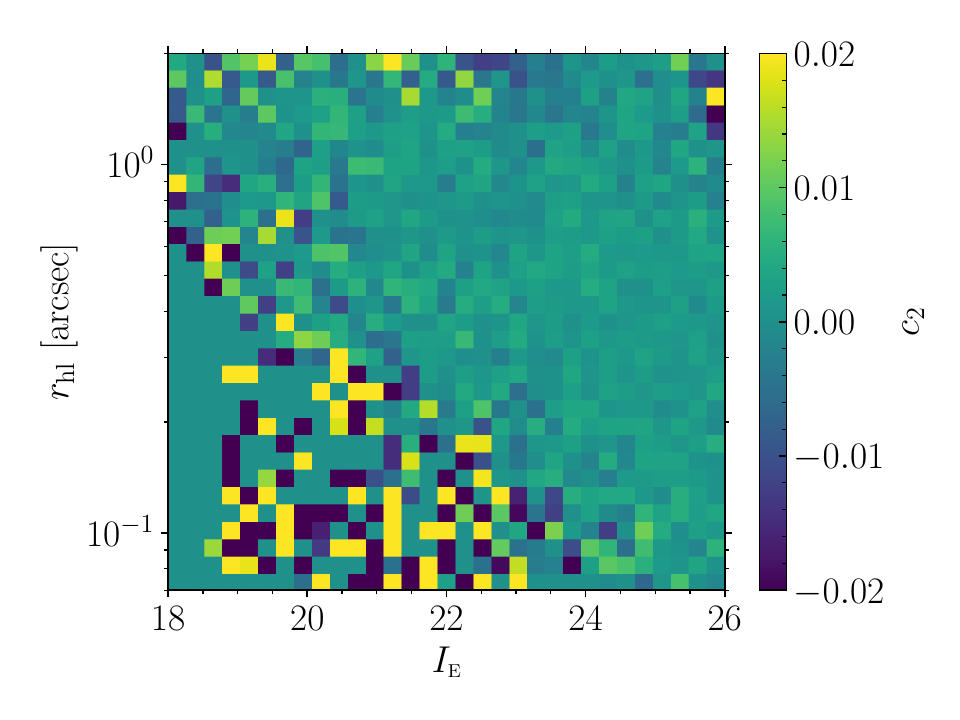}
\caption{Additive biases as functions of magnitude and half-light radius in the output selection.
Despite the noise in each bin due to the limited data volume, a clear trend in the map across magnitude and radius is visible.}
\label{fig:c_map}
\end{figure}

In the absence of spatial variability, it is still expected that biases should depend on galaxy brightness and size.
Therefore, we calculated the average correction as a function of observed magnitude and half-light radius, and then corrected the catalogue according to which bin the galaxy fell in.
Figure\;\ref{fig:c_map} shows the estimated $c_1$ and $c_2$ as functions of $\IE$ and $r_\text{e}$ that were used in correction of the catalogue.

\begin{figure}
\centering
\includegraphics[width=\columnwidth]{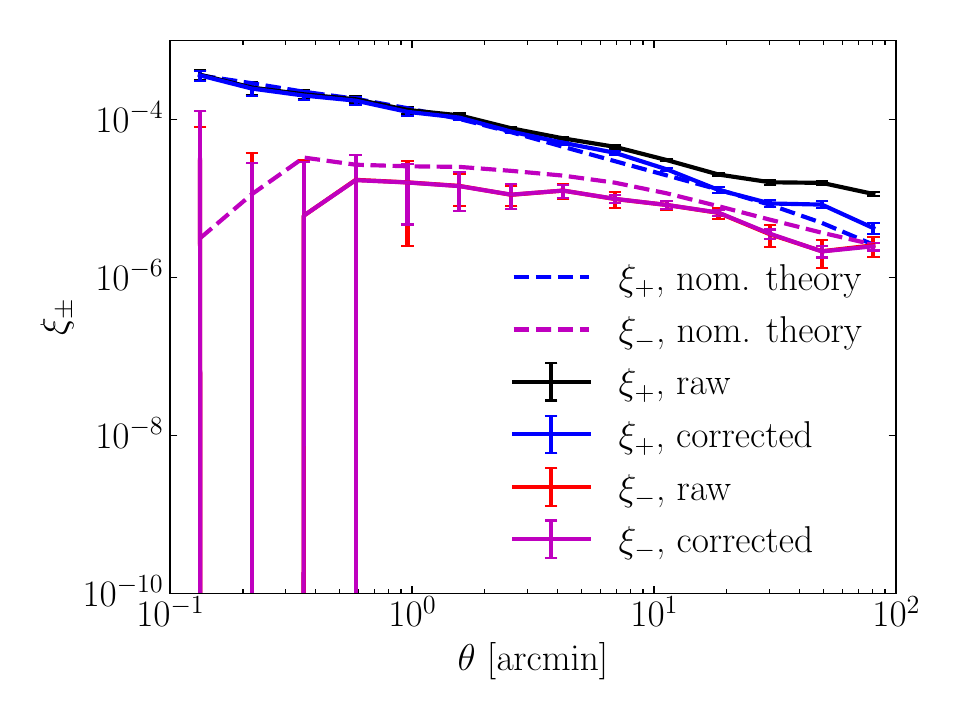}
\caption{Two-point correlation functions estimated on raw and corrected catalogues after a lensing sample cut (see Sect.\;\ref{sec:bias} for details).
A reduction in power due to the correction at scales of over \ang{;10} is evident.
However, excess power in $\xi_-$ can still be observed, which is likely due to the adoption of a raw, uncalibrated redshift distribution, the assumed nominal cosmology, or non-linear physics.}
\label{fig:2pcf}
\end{figure}

Figure\;\ref{fig:2pcf} shows a qualitative comparison between the raw and corrected catalogue performed with two-point correlation functions, $\xi_+$ and $\xi_-$, on scales between \ang{;0.1} and \ang{;100}, estimated with \texttt{treecorr} \citep{jarvis2004}.
As an additional validation, we also calculated a nominal flat $\Lambda$-cold-dark-matter cosmological model curve \citep[$\Omega_\text{c}=0.265$, $\Omega_\text{b}=0.049$, $H_0=67.3\,\kmsMpc$, $n_\text{s}=0.966$, and $\sigma_8=0.812$, ][]{aghanim2020},
calculated with \texttt{pyccl} \citep{chisari2019}.
It is worth stressing that this is not a quantitative assessment but only a general indication of the overall quality of the measurement.
The main reason is that the theory curve assumes a nominal redshift distribution that was calculated with the available photometric redshifts, which were not calibrated.
However, we did apply a magnitude cut for $\IE<24.5$, as will be most likely the case for the primary cosmological analysis of \Euclid.
We also removed large objects (see Fig.\;\ref{fig:mag-size}) according to $r_\text{hl}>\ang{;;2}$.
Additionally, we applied photometric redshift quality cuts via \texttt{PHZ\_FLAGS}=0 or 12.
In this context, given all the limitations and caveats of our analysis (which include no calibration of the redshift distribution or tuning of the theory curve), it is not surprising to see deviations between data and theory, which is evident for $\xi_-$.
Incidentally, the data and model for $\xi_+$ appear consistent, while there are deviations observed in $\xi_-$. This could be caused by the raw, uncalibrated redshifts, the assumed nominal cosmology, or even the modelling of non-linear physics.
Therefore, this comparison should be considered just indicative and definitely not optimal for a cosmological analysis. 
In summary, $\xi_+$ and $\xi_-$ broadly recover the expected trends, barring the excess power in the latter, which likely reflects the caveats discussed above.
A similar analysis carried out with aperture mass statistics \citep{schneider1996} also shows that the $B$ modes are an order of magnitude smaller than the $E$ modes.

\section{Cluster lensing profiles}\label{sec:profiles}


In this section, we proceed by assessing the quality of the measurement with self-consistency tests and cross-checks with external data.
All systematic metrics are defined in terms of the comoving radius from the centre of the cluster, $R_\text{c}$.\footnote{The comoving radius is defined as $R_\text{c}=(1+z)\,\theta\,D_\text{a}(z)$, where $z$ is the redshift, $\theta$ is the angular scale, and $D_\text{a}(z)$ is the angular diameter distance, which depends on cosmology.}
The conversion between angles and $R_\text{c}$ is cosmology dependent.
Here, we assumed a nominal flat cosmological geometry ($\Omega_\text{m}=0.3$ and $H_0=70\,\kmsMpc$)
in order to convert angular scales to comoving distances.
It is fair to ask how sensitive the shear profiles are to the assumption of a fixed cosmology.
In general, small differences in the assumed cosmology would introduce a slight rescaling of all the distances and a correlation between shear profile bins, both of which are irrelevant for the Q1 survey area, but might be for DR1.

\begin{figure}
\centering
\includegraphics[width=\columnwidth]{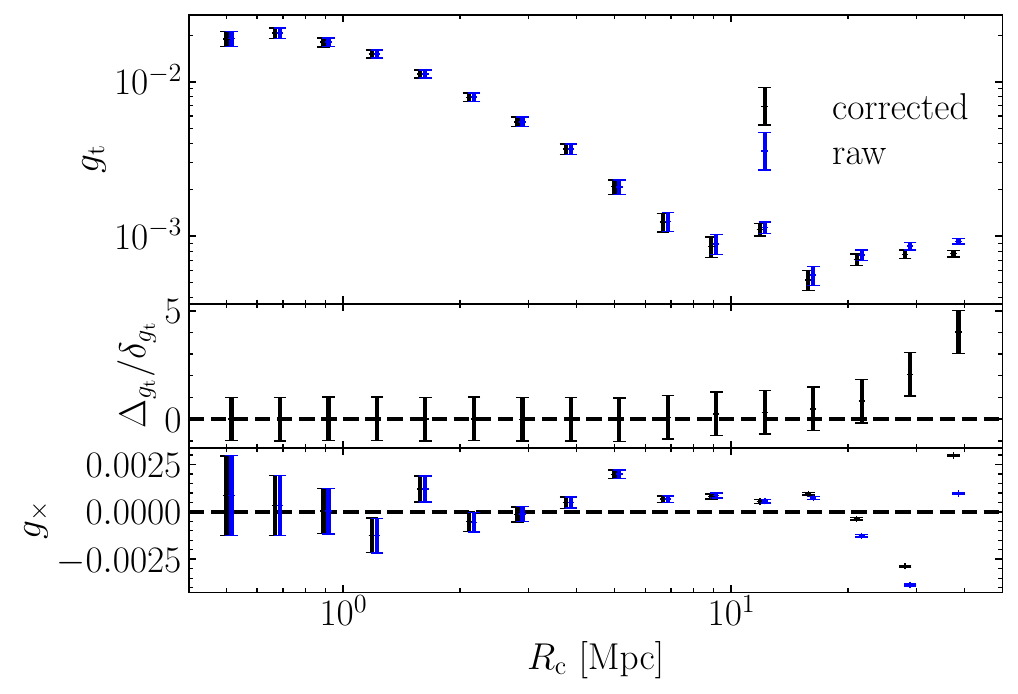}
\caption{Average lensing profile of all MaDCoWS2 cluster candidates detected in the Q1 fields, with the raw and corrected catalogues.
Only shape noise uncertainties are shown. \emph{Top panel:} Tangential shear profile.
\emph{Middle:} Normalised difference between raw and corrected catalogues in units of standard deviation.
\emph{Bottom:} Cross-shear profile.
The profiles are consistent up to a comoving radius of 20\,Mpc where a deviation between raw and corrected catalogues of almost 5 standard deviations starts to show up.}
\label{fig:madcows_corrected_vs_raw}
\end{figure}

Figure\;\ref{fig:madcows_corrected_vs_raw} shows the reduced tangential and cross shear of galaxies measured around the Massive and Distant Clusters of WISE Survey 2 (MaDCoWS2) candidates without any cuts in purity or cluster member decontamination \citep{thongkham2024}.
There are 495 MaDCoWS2 candidates in Euclid Q1, with 163 detections in EDF-F and 332 in EDF-S.
The sample covers nearly 10\;Gyr of cluster evolution history out to $z \approx 2$.
The cluster lensing analysis follows \citet{EP-Sereno} and \citet{ser25_self}, which we refer to for further details.
The lensing of MaDCoWS2 clusters in Q1, including a detailed discussion about the background selection, is the subject of a dedicated paper (Sereno et al, in prep.).

Additionally, Fig.\;\ref{fig:madcows_corrected_vs_raw} demonstrates that the profiles corresponding to the raw and corrected catalogues are fully consistent up to a comoving radius of about 20\;Mpc, where a deviation of almost 5 standard deviations starts to show up.
This is somehow consistent with a similar observation on $\xi_+$ above $\ang{;10}$ (see Fig.\;\ref{fig:2pcf}).
Nonetheless, the cross-shear, which is a diagnostic for systematic errors, appears broadly consistent with zero and no clear trend can be observed.
However, the statistical uncertainty associated with the data point appears underestimated relative to the statistical scatter, especially on scales above 10\;Mpc.
It is worth observing that both the raw and corrected catalogues provide consistent estimates of the inner regions of the clusters.
However, the improvement of the correction is clearer on larger scales, where the PSF modelling incompleteness may be more relevant.
In our separate paper, we calculated a $p$-value of $4\times10^{-3}$ across all the bins, which would not be sufficient to reject the null hypothesis at a $3\,\sigma$-confidence level.
However, this $p$-value is dominated by clusters at redshift $z>1$, which may be more affected by systematic errors.
In the remainder of this section, we will only consider the corrected catalogue.

\begin{figure}
\centering
\includegraphics[width=\columnwidth]{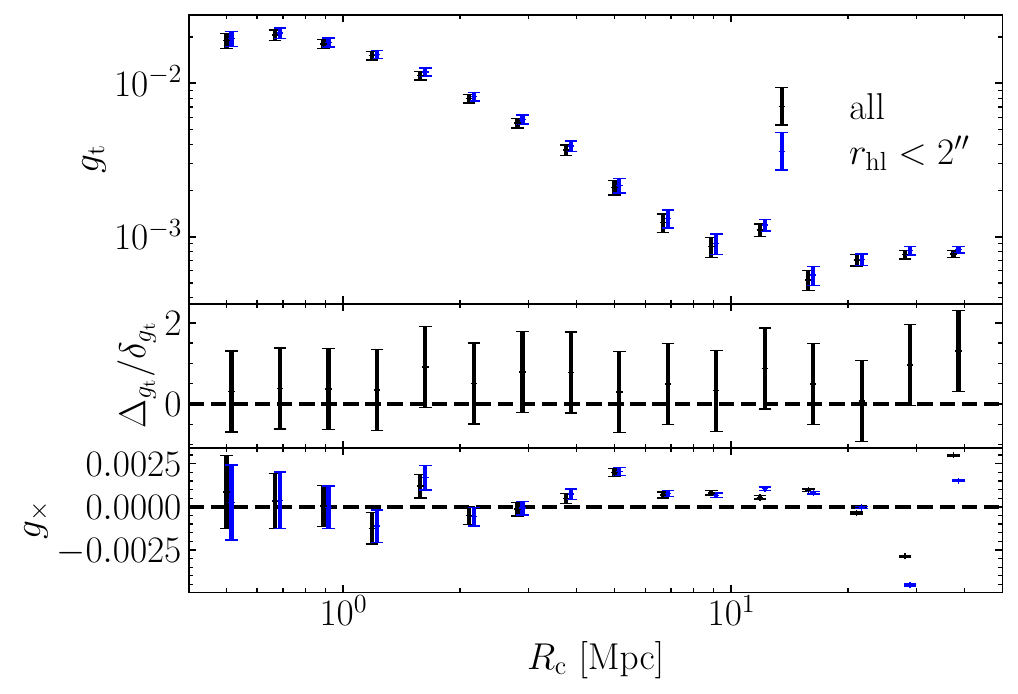}
\caption{%
Same as Fig.\;\ref{fig:madcows_corrected_vs_raw}, but comparing corrected shapes (`all') to the case after cutting in half-light radius ($r_\textrm{hl} < 2\arcsec$).
The profile of the full catalogue is, on average, about half a standard deviation higher compared to after the cut.}
\label{fig:madcows_large_sizes}
\end{figure}

As noted for the magnitude--size distribution of Fig.\;\ref{fig:mag-size} and the two-point correlation functions of Fig.\;\ref{fig:2pcf}, the reportedly-large galaxies may still impact the lensing analysis presented here.
Figure\;\ref{fig:madcows_large_sizes} quantitatively shows that the cut would reduce the profile by half a standard deviation, which may be indicative of a positive multiplicative bias due to those objects left in the full catalogue.
Although the impact of this cut can be ignored for the current analysis, we note that it might still be relevant for the full processing of DR1.

Furthermore, we cross-validated our measurements with a fully independent catalogue, namely the Dark Energy Survey (DES) year-3 measurements based on the \textsc{Metacalibration} method \citep{gatti2021}.
In general, direct comparisons between catalogues obtained with different image resolutions, bias corrections, band passes, or depth might be misleading, as the spatial distribution of light emission can differ \citep{schrabback2018}.
Most importantly, due to the larger ground-based PSF and lower depth, galaxies observed by DES will generally appear much larger and fainter than for \Euclid.
However, the estimated reduced tangential shear profile is still expected to be consistent when a matched source catalogue is used \citep{EP-Sereno}.
Therefore, any observed differences in shear signal should most likely be due to differential calibration errors.

Because the DES catalogue is shallower than \lensmc by about two magnitudes compared to Fig.\;\ref{fig:count}, we considered cluster lenses detected by DES, which maximises the signal and any relevant differences between the catalogues.
We used the sample of galaxy cluster candidates detected in the DES year-1 photometric data by the red-sequence matched-filter probabilistic percolation cluster finding algorithm \citep[\texttt{redMaPPer},][]{rykoff2014,rykoff2016,mcclintock2019}. 
We found 139 clusters in EDF-F and EDF-S in the redshift range $0.21 < z_\text{d} < 0.86$ and with richness $20 < \lambda < 175$. 
As lensed sources, we considered galaxies with $0.2 < z_\text{p} < 1.2$ and $z_\text{d} > z_\text{d} +0.1~(1+ z_\text{d})$, where $z_\text{p}$ is the DES point estimate for redshift with the \texttt{BPZ} method and $z_\text{d}$ is the lens (deflector) redshift.
The source selection is based on the photo-$z$ of the DES survey, which we took as a reference for both catalogues.

\begin{figure}
\centering
\includegraphics[width=\columnwidth]{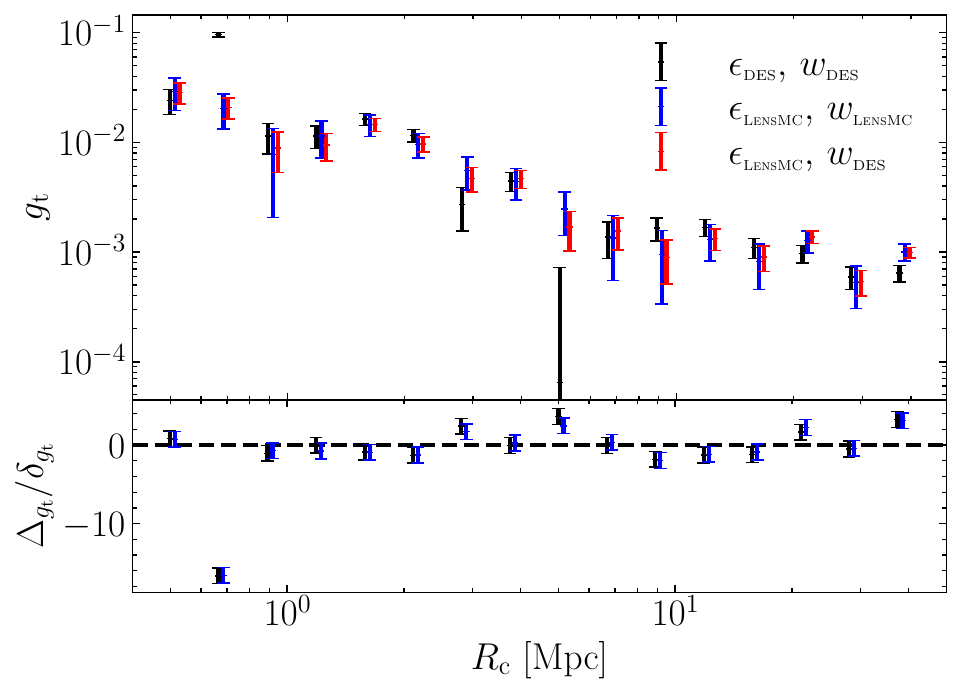}
\caption{Average shear profiles of \texttt{redMaPPer} clusters detected in the DES year-3 fields covered by Q1,
using different combinations of DES/\lensmc ellipticity/weights.
\emph{Top panel:} Tangential shear.
\emph{Bottom:} Normalised fractional difference between \lensmc (with \lensmc or DES weights) and DES.
Overall, the DES and \lensmc measurements agree very well, except for the outlier at 0.65\;Mpc (see text for details).}
\label{fig:redmapper_des_vs_lensmc}
\end{figure}

Figure\;\ref{fig:redmapper_des_vs_lensmc} shows the tangential shear profiles obtained for three combinations of shapes and weights from the two matched samples: (i) DES ellipticities with DES weights; (ii) \lensmc ellipticities with DES weights; (iii) \lensmc ellipticities with \lensmc weights.
This comparison is meant to assess any impact on results due to systematic differences between the two shear catalogues.
Despite the fundamental differences in survey imaging and shear measurement methodology, the profiles are consistent, except for one particular bin of the DES sample at 0.65\;Mpc.
However, it is worth noting that this outlier disappears if we consider the excess surface density, where weights account for lens-source separation.
Therefore, this spurious outlier might be just due to a fraction of sources with overestimated shear or weight in the DES catalogue.
We quantified any potential discrepancy between the two measurements with differential multiplicative, $\delta m$, and additive, $\delta c$, biases.
Because the two samples are matched and lenses and sources are the same, any difference between the catalogues cannot be explained in terms of correlations in the large-scale structure.
In fact, the two catalogues must show a relatively high degree of correlation because they effectively contain the same objects.
For this reason, we only included the uncertainty of the reference dataset (DES), and fitted the difference between \lensmc and DES.
After the removal of the outlier, we measured $\delta m = (-5 \pm 8)\times10^{-2}$ and $\delta c = (2 \pm 1) \times 10^{-4}$ over the full range of distances, which is consistent with no bias.
However, most of the mild tension is driven by the distant bins, and we found $\delta m = (2\pm 8)\times10^{-2}$ and $\delta c = (-0.6 \pm 1) \times 10^{-4}$ after the removal of the last bin, and $\delta m = (-1 \pm 8)\times10^{-2}$ and $\delta c = (-0.5 \pm 2) \times 10^{-4}$ after the removal of the last two bins.

\begin{figure}
\centering
\includegraphics[width=0.8\columnwidth]{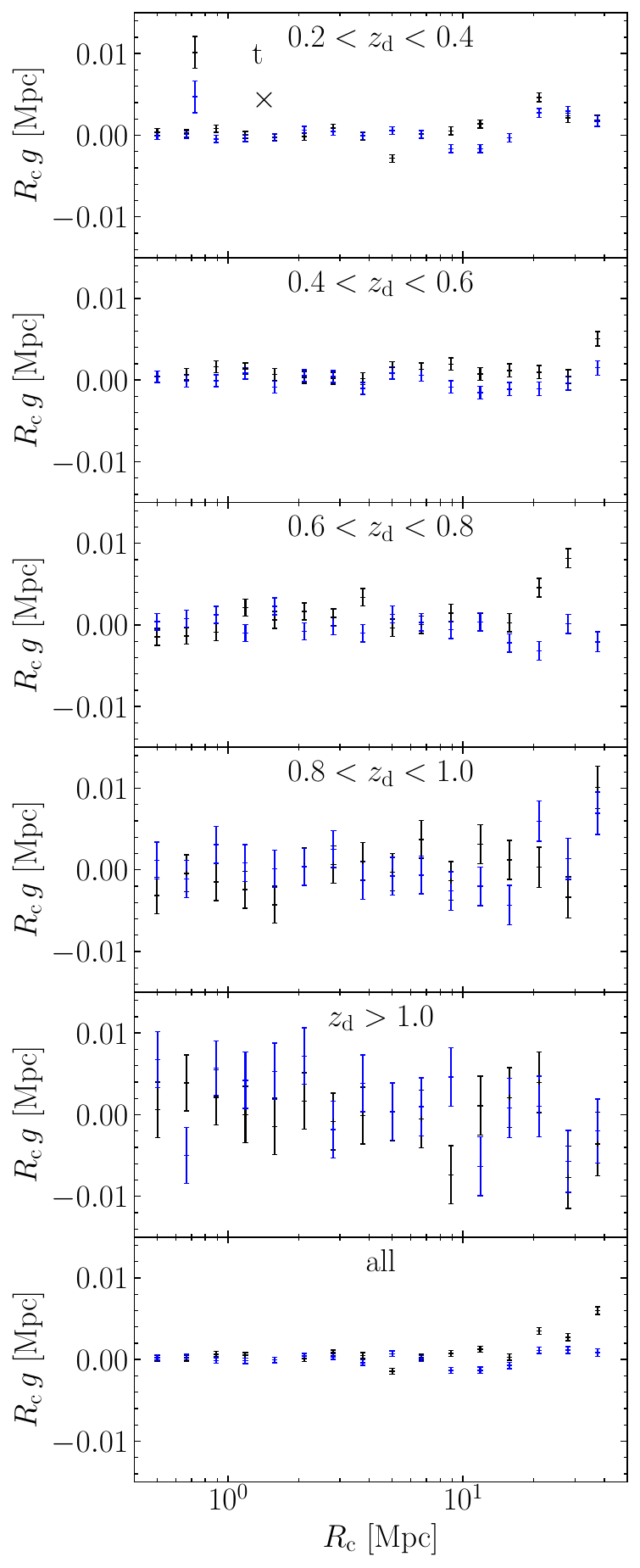} 
\caption{Average tangential and cross-shear profiles of \num{4000} MaDCoWS2-like random clusters across Q1.
Each panel shows a defined bin of lens redshifts, including a high-redshift bin, and the total.
The profiles are consistent with zero, except for the low-redshift, large radii bins (see Sect.\;\ref{sec:profiles} for details).}
\label{fig:madcows_random_clusters}
\end{figure}

Residual PSF or FoV-dependent biases could still imprint a non-null signal around random pointings across the footprint.
We simulated a MaDCoWS2-like random catalogue of \num{4000} clusters where redshift and $\text{S/N}$ values were assigned by bootstrap resampling the original catalogue.
Each position of a lens in the unmasked area was simulated by randomly drawing a source from the shear catalogue, and randomly moving it in a window of $\pm 10\arcsecond$ in celestial coordinates.
Although a full analysis of the MaDCoWS2 clusters was performed in our separate paper, for this validation test we only considered shape noise as a proxy for the actual statistical scatter.
Figure\;\ref{fig:madcows_random_clusters} shows the tangential and cross-shear components in four redshift bins of the lenses between 0.2 and 1, one bin for lenses beyond redshift of 1, and one cumulative bin for all lens redshifts.
The shear profiles are consistent with the null signal in most of the radial bins, except for the large radii at low redshift, which suggests the possibility of residual systematic errors most likely due to border effects, incomplete azimuthal coverage, or PSF residuals on those scales.
The deviation is driven primarily by low-redshift lenses, whose angular extension on the sky is much larger than high-redshift lenses where the effect is negligible.

\section{Conclusions}\label{sec:conclusions}

We presented an analysis of the \Euclid images published as part of the Q1 release and produced a shear catalogue of stacked images with the \lensmc method officially adopted for DR1.
Although the area covered is relatively small, this provided a good test in anticipation of DR1.
We measured galaxies to full depth and included all of them without any selection cuts in all our validation tests, except for those cuts required at the detection stage.
Our tests include characterising any FoV-dependent effects, bearing in mind that the effective FoV of our analysis is defined by the stacked mosaic.
Additionally, we developed an empirical additive bias correction based on morphological binning of galaxy properties (magnitude and size), which directly correlate with observational systematic effects (e.g., the PSF).
In contrast to the FoV-level characterisation, which does not show significant trends (except for a constant offset), the morphological characterisation does show patterns.
However, because of the limited data volume, both approaches tend to be noisy.
Although we expect that our correction will not be accurate enough to meet the cosmic shear requirements of the full Euclid Wide Survey analysis, we verified that our bias correction indeed suppresses lensing power at scales above \ang{;10}, and broadly recovers the expected theory curve.

Since our goal is the application to cluster lensing, we focussed on validating our measurements directly with stacked lensing profiles.
We proceeded by estimating the tangential and cross-shear profiles around MaDCoWS2 clusters identified in the two fields in the south, finding good consistency and a clear suppression of power at a transverse radial scale of 20\;Mpc.
Additionally, we quantified the impact of large spurious galaxies on the cluster profiles, finding it to be at about half a standard deviation on average across the radial bins.
Furthermore, we cross-matched our sources with the DES year-3 catalogue in the same area and calculated the shear profiles of \texttt{redMaPPer} clusters with different combinations of shape and weights, overall finding very good consistency across the two methodologies, except for a single outlier in the DES sample, which disappears in the full lensing cluster analysis.
Finally, we simulated MaDCoWS2-like random pointings across the Q1 footprint, and found that the measurement is consistent with the null signal, as expected, except for comoving radii above 10\,Mpc at low redshift.
One potential explanation is that low-redshift clusters appear much larger than high-redshift ones, and therefore edge effects due to the limited area or residual PSF errors may still be impacting the measurements.

In summary, with all the caveats and limitations of our analysis, we demonstrated that we can successfully measure galaxies to full depth with a surface number density of $75\;\unit{arcmin^{-2}}$ for $\IE<27$, and use all galaxies without additional cuts to measure the cluster lensing profiles out to redshift $z>1$ and up to comoving radii of $20\;\unit{Mpc}$ with very good control of the systematic errors.
Beyond these scales, the cross-component of the lensing profiles becomes significant, suggesting that other effects may be contributing.
One possibility is the impact of systematic errors, as discussed above.
Another possibility, more connected to the physics of clusters and their interaction with the local environment, could be 
the gravitational interaction between clusters and filaments, leading to non-spherical symmetry and generally the non-equilibrium of the mass distribution, particularly beyond the splashback radius.

To conclude, the main result of our work, which to our knowledge would be unprecedented in the field, is that the quality of \Euclid data and our control of systematic errors allow us to constrain the lensing profiles of clusters with relatively low masses of $10^{14}M_\odot$ out to $z\approx2$ and up to a comoving radius of $20\,\unit{Mpc}$ over nearly 10\;Gyr of evolution history.
Furthermore, a $\Lambda$CDM model can be fitted to these profiles on small and larger radii, including the unprecedented possibility of constraining the 2-halo term from our measurements.
We refer the interested reader to the main lensing analysis presented in a separate paper (Sereno et al., in prep.).
The processed galaxy catalogue is available upon request.

%
%

\begin{acknowledgements}
GC acknowledges support provided by the Joint Research Advancement Programme between Nagoya University and University of Edinburgh.
GC thanks the United Kingdom Space Agency for additional support.
GC acknowledges the use of the IRIS infrastructure and \Euclid clusters in Edinburgh to perform the analysis of the Q1 data.
GC thanks N.~Hambley for help on \emph{Gaia} data.
MS acknowledges financial contributions from: contract INAF mainstream project 1.05.01.86.10; INAF Theory Grant 2023: Gravitational lensing detection of matter distribution at galaxy cluster boundaries and beyond (1.05.23.06.17); INAF Guest Observer Grant 2024: Towards anchoring the mass scale of galaxy clusters with galaxy kinematics (1.05.24.02.15); and contract Prin-MUR 2022 supported by Next Generation EU (no.\ 20227RNLY3: The concordance cosmological model: stress-tests with galaxy clusters).
HM is supported by JSPS Kakenhi Grant Numbers: JP22K21349, JP23H00108, and JP24KK0065.

Some of the results in this paper were derived using the \texttt{healpy} and \healpix packages (\href{https://healpix.sourceforge.io}{healpix.sourceforge.io,}).
Figure\;\ref{fig:footprint} made use of \emph{Gaia} DR3 data (\url{https://doi.org/10.17876/gaia/dr.3/1}) from the European Space Agency (ESA) mission \emph{Gaia} (\url{https://www.cosmos.esa.int/gaia}), processed by the Gaia Data Processing and Analysis Consortium (DPAC, \url{https://www.cosmos.esa.int/web/gaia/dpac/consortium}).

\AckQone

\AckEC

\end{acknowledgements}

%
%

\bibliography{Euclid, Q1, references} 

%

  

\end{document}